\documentclass[twocolumn,
notitlepage,nofootinbib,preprintnumbers,amssymb,superscriptaddress]{revtex4-1}
\usepackage{amsfonts,amssymb,amsmath,graphicx,color,bm}
\definecolor{ultramarine}{rgb}{0.07, 0.04, 0.56}
\definecolor{cadmiumgreen}{rgb}{0.0, 0.42, 0.24}
\definecolor{indigo(dye)}{rgb}{0.0, 0.25, 0.42}
\usepackage[linktocpage=true]{hyperref}
\hypersetup{
colorlinks=true,
citecolor=red, 
linkcolor=red, 
urlcolor=indigo(dye),
}

\newcommand{\be}{\begin{equation}}  
\newcommand{\ee}{\end{equation}}
\newcommand{\bem}{\begin{bmatrix}}
\newcommand{\eem}{\end{bmatrix}}

\newcommand{\mR}{\mathcal{R}}

\allowdisplaybreaks

\begin{document}

\title{
Spontaneous vectorization in the presence of vector field coupling to matter
}

\author{Masato Minamitsuji}
\affiliation{Centro de Astrof\'{\i}sica e Gravita\c c\~ao  - CENTRA, Departamento de F\'{\i}sica, Instituto Superior T\'ecnico - IST, Universidade de Lisboa - UL, Av. Rovisco Pais 1, 1049-001 Lisboa, Portugal}

\begin{abstract}
We examine the possibility of spontaneous vectorization in the vector-tensor theories with the vector conformal and disformal couplings to matter. We study the static and spherically symmetric solutions of the relativistic stars with the nontrivial profile of the vector field satisfying the boundary conditions $A_\mu\to 0$ at the spatial infinity, where $A_\mu$ represents the vector field. First, we study the linear perturbations about the general relativistic (GR) stellar solutions with the vanishing vector field $A_\mu =0$. We show that the pure vector disformal coupling causes the ghost or gradient instability of the GR stars, indicating the breakdown of the hyperbolicity on the GR stellar backgrounds. On the other hand, the pure conformal coupling causes the tachyonic instability of the GR solutions and would lead to the spontaneous growth of the vector field toward the nontrivial solutions as in the manner of spontaneous scalarization in the scalar-tensor theories. We then construct the static and spherically symmetric solutions of the relativistic stars with $0$ and $1$ nodes of the vector field in the presence of the pure disformal coupling. We find that the properties of the solutions with the nontrivial vector field are similar to those of the generalized Proca theories. As in the mass-radius diagram the branches of the $0$ and $1$ node solutions are disconnected to that of the GR solutions in the lower density regimes, they may be formed from the selected choice of the initial conditions. We also construct the $0$ node solutions in the case of the pure conformal coupling, and show that the branch of these solutions is connected to that of the GR  solutions in both the low and high density regions and hence would arise spontaneously via the continuous evolution from the GR solutions. Finally, we briefly discuss the combined effects of the conformal and disformal couplings on the nontrivial solutions.
\end{abstract}

\maketitle

\section{Introduction}
\label{sec1}

It is well-known that 
general relativity (GR) has passed all experimental tests
from the Solar System to the strong-field and low-velocity regime of binary pulsars~\cite{Will:2014kxa}.
With the dawn of gravitational-wave astronomy,
a new frontier for testing GR in strong-field and high-velocity regimes has opened
\cite{Berti:2015itd,Berti:2018vdi,Berti:2018cxi}.
Some of the scalar-tensor theories of gravitation
not only pass the Solar System tests, 
but
also allow for large deviations from GR 
in the relativistic star interiors such as neutron stars,
through the process called spontaneous scalarization~\cite{Damour:1993hw}.
Spontaneous scalarization would be triggered
by the tachyonic growth of the scalar field inside the relativistic stars.
In the simplest case
where the metrics in the Jordan and Einstein frames 
are related by the conformal transformation,
${\tilde g}_{\mu\nu}= A(\phi)^2 g_{\mu\nu}$,
the GR solution with the vanishing scalar field $\phi=0$
suffers from the tachyonic instability if $(\ln A)_{,\phi\phi}<0$, 
leading to the relativistic stars
with the nontrivial profile of the scalar field $\phi=\phi(x^\mu)$
\cite{Damour:1996ke,Harada:1997mr,Harada:1998ge,Novak:1998rk,Palenzuela:2013hsa,Sampson:2014qqa,Pani:2014jra,Silva:2014fca}.
For the exponential coupling $A(\phi)=\exp(\gamma_\alpha \phi^2/2)$,
spontaneous scalarization of the static and spherically symmetric stars
happens for $\gamma_\alpha\lesssim -4.35$ \cite{Harada:1997mr,Harada:1998ge},
depending weakly on the equation of state 
and fluid properties~\cite{Novak:1998rk,Silva:2014fca}.

An interesting question is
whether phenomena similar to spontaneous scalarization 
can be induced by another field,
for instance, the vector field.
The possibility of spontaneous vectorization
triggered by the tachyonic instability of the vector field $A_\mu$
has been recently studied
in Refs. \cite{Ramazanoglu:2017xbl,Ramazanoglu:2019gbz,Annulli:2019fzq,Ramazanoglu:2019jrr,Kase:2020yhw}.
The authors of Ref.~\cite{Annulli:2019fzq}
have studied the model 
of the vector-tensor theory with the nonminimal couplings to the Ricci tensor $\beta R^{\mu\nu}A_\mu A_\nu$
(as well as $\gamma R A^\mu A_\mu$),
and found that below the threshould couplings $\beta={\cal O}(-1.0)$
the GR solutions with the vanishing vector field $A_\mu=0$ suffer from the tachyonic instability.
They have also constructed the 
static and spherically symmetric solutions with the nontrivial vector field profile with $0$ nodes
for $\beta={\cal O} (-0.1)$,
which would be formed from the selected choice of the initial conditions, 
rather than from the tachyonic instability of the GR solutions.
A similar model in the context of the generalized Proca theories \cite{Heisenberg:2014rta,DeFelice:2016cri}
with the nonminimal coupling to the Einstein tensor $G^{\mu\nu}A_\mu A_\nu$,
which is free from the Ostrogradsky instability,
has been studied in Ref.~\cite{Kase:2020yhw}.
The $0$ node solutions have been shown to exhibit the similar behavior to those in the above model,
and the $1$ node solutions have also been studied.
While the coupling parameter for the existence of the $1$ node solutions
is consistent with the instability of the GR solutions,
in the mass-radius diagram 
the branch of the nontrivial vector field solutions
is not smoothly connected 
to that of the GR  solutions with the vanishing vector field.
Thus, 
it may be formed from the selected choice of the initial conditions
as the $0$ node solutions.
These properties of the relativistic stars
were much different from those of spontaneous scalarization.

While these studies have analyzed the possibility of spontaneous vectorization
from the viewpoint of the Jordan frame
where the nonminimal coupling of the vector field to the spacetime curvature is present, 
the other approach
is to take the viewpoint of the Einstein frame
where the nonminimal coupling of the vector field to matter is present. 
Ref. \cite{Ramazanoglu:2017xbl} 
has shown that
the conformal coupling to the vector field ${\tilde g}_{\mu\nu}= C(X)g_{\mu\nu}$,
where 
\begin{eqnarray}
X:= g^{\mu\nu} A_\mu A_\nu,
\end{eqnarray}
is the norm of the vector field $A_\mu$
and $C(X)$ is the regular function of $X$,
could trigger the tachyonic instability of the vector field,
and 
the explicit stellar solutions with the nontrivial profile of the vector field 
have been constructed
for the coupling $C(X)=e^{\beta X}$ ($\beta<0$) and the nonzero mass of the vector field.
More recently, 
Ref. \cite{,Ramazanoglu:2019jrr} has also suggested
that the vector disformal coupling \cite{Kimura:2016rzw},
${\tilde g}_{\mu\nu}= g_{\mu\nu}+ B(X)A_\mu A_\nu$,
where $B(X)$ is also the regular function of $X$,
could also trigger the tachyonic instability of the vector field in the relativistic stars.
We note that the Jordan frame formulation of the case of the pure conformal coupling
was recently presented in Ref. \cite{Ramazanoglu:2019tyi}.
We also note that 
in the context different from spontaneous vectorization
hairy relativistic star solutions in generalized vector-tensor theories
have been recently studied in Refs.~\cite{Chagoya:2017fyl,Kase:2017egk}.

The purpose of the paper is 
to present the more quantitative study 
of the relativistic stars with the nontrivial profile of the vector field
in the presence of conformal and disformal couplings to matter,
and 
to address whether the solutions satisfying the boundary conditions $A_\mu=0$ at the spatial infinity 
can be the vectorized solutions of the relativistic stars.
In this paper, 
we consider the vector-tensor theory with the direct coupling of the vector field and matter,
whose action is given by 
\begin{eqnarray}
\label{action}
S
&=& \frac{1}{2\kappa^2} 
   \int d^4 x\sqrt{-g}
 \left[
  R-\frac{1}{4}F^{\mu\nu}F_{\mu\nu}
-\frac{1}{2}m^2A^\mu A_\mu
 \right]
\nonumber\\
&+&
\int d^4 x \sqrt{-{\tilde g}}{\cal L}_m ({\tilde g}_{\mu\nu},\Psi)
=:S_g+S_m,
\end{eqnarray}
where 
$\kappa^2=8\pi G/c^4$ with $G$ being the gravitational constant and $c$ being the speed of light,
$g_{\mu\nu}$ is the metric in the Einstein frame,
$R$ is the scalar curvature associated with $g_{\mu\nu}$,
$A_\mu$ is the vector field,
$F_{\mu\nu}:=\partial_\mu A_\nu-\partial_\nu A_\mu$ is the field strength,
and $m^2$ is the bare mass squared of the vector field.
We assume that 
the metric in the Jordan frame denoted by ${\tilde g}_{\mu\nu}$ 
is related to the Einstein frame one $g_{\mu\nu}$
by the combination of conformal and disformal couplings
\begin{eqnarray}
\label{disformal}
{\tilde g}_{\mu\nu}
=C(X)\left(g_{\mu\nu}+ B(X) A_\mu A_\nu\right).
\end{eqnarray}
The gravitational part of the action $S_g$ in Eq. \eqref{action} 
written in terms of the Jordan frame metric ${\tilde g}_{\mu\nu}$
belongs to a class of the beyond-generalized Proca theories \cite{Heisenberg:2016eld,Heisenberg:2016lux,Kimura:2016rzw}.
In this paper,
we will work in the units where the speed of light to be unity $c=1$,
unless it should be shown explicitly.

For the analysis of the solutions of the relativistic stars, 
we will employ only the polytrope equation of state~\cite{Damour:1993hw,Shapiro}
(See Eqs. \eqref{poly1} and \eqref{pressure}).
Since this equation of state would be less realistic,
we will not make any comparison with the observational data 
associated with the mass and radius of the neutron stars,
and focus on the physical properties of the solutions.
In Ref~\cite{Kase:2020yhw},
it has also been shown that 
the qualitative features of the relativistic stars in the generalized Proca theories
do not depend on the choice of the equations of state.
We expect 
that in our model
the properties of the star 
are also less sensitive to the choice of the equation of state.

The organization of the paper is as follows:
In Sec.~\ref{sec2},
we will derive the set of the covariant equations of motion
by varyng the action \eqref{action}
and analyze the stability of the GR solutions.
In Sec.~\ref{sec3},
we will apply the above formulation to the static and spherically symmetric spacetme,
reduce the set of the equations of motion to the modified Tolman-Oppenheimer-Volkov system,
and derive the interior and exterior solutions of the star.
In Sec.~\ref{sec4},
we will analyze the properties of the test vector field
on top of the constant density star,
and
estimate the typical values of the conformal and disformal couplings
to enhance the vector field at the center of the star.
In Sec.~\ref{sec5},
we will explicitly construct solutions of the relativistic star
with the nontrivial vector field profile 
in the simplest model of the pure disformal coupling
and discuss their properties.
In Sec.~\ref{sec6},
we will analyze the simplest model with the pure conformal coupling.
In Sec.~\ref{sec7},
we will briefly discuss the combined effects of 
the conformal and disformal couplings on the nontrivial solutions.
The last Sec.~\ref{sec8} 
will be devoted to a brief summary and conclusion.

\section{the vector-tensor theories and the GR solutions}
\label{sec2}

In this section,
we derive the covariant equations of motion
by varying the action \eqref{action}
with respect to
the Einstein frame metric $g_{\mu\nu}$ and the vector field $A_\mu$.

\subsection{The covariant equations of motion}
\label{sec21}

Varying the action \eqref{action} with Eq. \eqref{disformal}
with respect to the vector field $A_\nu$,
we obtain the equation of motion for the vector field
\begin{eqnarray}
\label{veq}
&&
 \nabla_\mu F^{\mu\nu}
-m^2 A^\nu
=
-2\kappa^2
\sqrt{1+BX}
C^3
\left[
B  {\tilde T}^{\nu\rho}  A_\rho
\right.
\nonumber\\
&&
\left.
+
\left(
 B_X 
 {\tilde T}^{\rho\sigma}
  A_\rho
  A_\sigma
+\frac{C_X}{C^2}
{\tilde T}
 \right)
A^\nu 
\right],
\end{eqnarray}
where 
$A^\mu:= g^{\mu\nu}A_\nu$,
and 
we have defined the energy-momentum tensor of matter 
 in the Jordan frame
\begin{eqnarray}
{\tilde T}^{\alpha\beta} 
=\frac{2}{\sqrt{-\tilde g}}
\frac{\delta (\sqrt{-{\tilde g}} {\cal L}_m) }
       {\delta {\tilde g}_{\alpha\beta}},
\end{eqnarray}
with 
the inverse and determinant of 
the metric tensor in the Jordan frame Eq. \eqref{disformal},
respectively,
given by Eq. \eqref{app1}.
On the other hand,
varying the action \eqref{action} with respect to the Einstein frame metric $g_{\mu\nu}$,
we obtain the gravitational equations of motion in the Einstein frame
\begin{eqnarray}
\label{eeq}
G^{\mu\nu}
=\kappa^2 
\left(
T^{(A) \mu\nu}
+T^{\mu\nu}
\right),
\end{eqnarray}
where we have defined 
the energy-momentum tensor of the vector field and matter in the Einstein frame,
respectively by
\begin{eqnarray}
\label{vector_em}
\kappa^2
T_{(A)}^{\mu\nu}
&=&\frac{1}{2}
 \left(
  F^{\mu\alpha}F^\nu{}_\alpha
-\frac{1}{4}g^{\mu\nu}F^{\rho\sigma}F_{\rho\sigma}
 \right)
\nonumber\\
&+&
\frac{m^2}{2}
  \left(
   A^\mu A^\nu
   -\frac{1}{2}g^{\mu\nu} X
  \right),
\end{eqnarray}
and
\begin{eqnarray}
\label{matter}
T^{\mu\nu}
&=&
\frac{2}{\sqrt{-g}}
\frac{\delta (\sqrt{-{\tilde g}} {\cal L}_m) }
       {\delta {g}_{\mu \nu}}
=
\sqrt{\frac{\tilde g}{g}}
\frac{\partial {\tilde g}_{\alpha\beta} }{\partial g_{\mu\nu}}
\left[
\frac{2}{\sqrt{-\tilde g}}
\frac{\delta (\sqrt{-{\tilde g}} {\cal L}_m) }
       {\delta {\tilde g}_{\alpha\beta}}
\right]
\nonumber\\
&=&
\sqrt{1+BX}C^3
\nonumber\\
&\times&
\left[
{\tilde T}^{\mu\nu}
+
A^\mu A^\nu
\left(
B_X
{\tilde T}^{\rho\sigma}
A_\rho A_\sigma
+\frac{C_X}{C^2} {\tilde T}
\right)
\right],
\end{eqnarray}
with ${\tilde T}:={\tilde g}_{\alpha\beta}{\tilde T}^{\alpha\beta}$,
$B_X:=\partial_X B$, 
and $C_X:= \partial_X C$.
Acting $\nabla_\nu$ on Eq. \eqref{eeq},
with the use of the contracted Bianchi identity $\nabla_\mu G^{\mu\nu}=0$,
we find
\begin{eqnarray}
\nabla_\mu T_{(A)}^{\mu\nu}
=-\nabla_\mu T^{\mu\nu}.
\end{eqnarray}
which 
after some algebra
leads to 
\begin{widetext}
\begin{eqnarray}
\label{ceq}
\nabla_\alpha {\tilde T}^{\alpha}{}_\nu 
&=&
-\frac{\nabla_\alpha\left(\sqrt{1+BX} C^2\right)}  
        {\sqrt{1+BX}C^2}
 {\tilde T}^\alpha{}_\nu
+
C
\left[
B  {\tilde T}^{\alpha \rho}  A_\rho
+\left(
  B_X {\tilde T}^{\rho\sigma}
  A_\rho
  A_\sigma
+\frac{C_X}{C^2}
{\tilde T}
\right)
A^\alpha
\right]
\nabla_\nu A_\alpha
\nonumber\\
&-&
\frac{2}
      {\sqrt{1+BX}C^2}
\nabla_\alpha
\left[
\sqrt{1+BX}
C^3
\left(
B_X {\tilde T}^{\rho\sigma}
  A_\rho
  A_\sigma
+\frac{C_X}{C^2}
{\tilde T}
\right)
A^\alpha
A_\nu
\right].
\end{eqnarray}
\end{widetext}

\subsection{The GR solutions and their stability}
\label{sec22}

The metric $g_{\mu\nu}$ satisfying the Einstein equations in GR
\begin{eqnarray}
\label{gr_eq}
G_{\mu\nu}=\kappa^2 T_{\mu\nu},
\end{eqnarray} 
and the vanishing vector field $A_\mu=0$ (and hence $X=0$)
are the solution of the theory \eqref{action} with Eq. \eqref{disformal}
for the coupling functions $B(X)$ and $C(X)$ are regular at $X=0$
and $C(0)=1$,
which can be written in terms of the Taylor series with respect to $X=0$,
\begin{eqnarray}
\label{regular_coupling}
C(X)&=&1+ C_1 X+\frac{1}{2}C_2 X^2
+\sum_{n=3} \frac{1}{n!}C_n X^n,
\\
\label{regular_coupling2}
B(X)&=& B_0 +B_1 X+\frac{1}{2}B_2X^2
+\sum_{n=3} \frac{1}{n!}B_n X^n,
\end{eqnarray}
where $C_n$ ($n=1,2,3,\cdots$) and $B_n$ ($n=0,1,2,\cdots$) are constants.
For the couplings \eqref{regular_coupling} and \eqref{regular_coupling2} regular at $X=0$,
substituting $A_\mu=0$ into Eqs. \eqref{disformal}, \eqref{vector_em}, and \eqref{matter},
we find that
\begin{eqnarray}
{\tilde g}_{\mu\nu}=g_{\mu\nu},
\quad
T_{\mu\nu}= {\tilde T}_{\mu\nu},
\quad
T^{(A)}_{\mu\nu}=0,
\end{eqnarray}
Eq. \eqref{eeq} reduces to the ordinary Einstein equation \eqref{gr_eq}
and Eq. \eqref{veq} is trivially satisfied.
This is the GR solution.

We then consider the small perturbations about the GR solutions with $A_\mu=0$.
At the level of the linear perturbations, 
the equations for the metric and vector field perturbations are decoupled.
Thus, the metric perturbation ${\delta g}_{\mu\nu}$ obeys
the perturbed Einstein equation in GR, 
\begin{eqnarray}
\delta G_{\mu\nu}=\kappa^2 \delta T_{\mu\nu},
\end{eqnarray}
where
$\delta G_{\mu\nu}$ and $\delta T_{\mu\nu}$
are the perturbed Einstein and energy-momentum tensor of matter,
respectively.
From Eq. \eqref{veq},
the vector field perturbation $\delta A^\mu$
satisfies 
\begin{eqnarray}
\label{pert_vec}
\nabla_\mu \delta F^{\mu\nu}
-m^2 \delta A^\nu
+2\kappa^2
\left[
B_0  {T}^{\nu\rho} \delta A_\rho
+C_1 {T}
\delta A^\nu 
\right]
=0,
\nonumber\\
\end{eqnarray} 
where $\delta F_{\mu\nu}:=\partial_\mu \delta A_\nu - \partial_\nu\delta A_\mu$
is the perturbed field strength.

We assume the perfect fluid form of the matter energy-momentum tensor
$T_{\mu\nu} =\left(\rho+p \right)u_\mu u_\nu+p g_{\mu\nu}$,
where $u_{\mu}$ is the four-velocity satisfying $g^{\mu\nu}u_\mu u_\nu=-1$.
We also assume a general static and spherically symmetric solution 
whose metric is given by 
\begin{eqnarray}
\label{sss}
ds^2
=g_{\mu\nu}dx^\mu dx^\nu
=-e^{\nu(r)}dt^2+e^{\lambda(r)}dr^2 +r^2 \gamma_{ab} dx^a dx^b,
\nonumber\\
\end{eqnarray}
where 
$\nu$ and $\lambda$ are functions of $r$,
$\gamma_{ab}$ is the metric of the unit two-sphere,
and 
the indices $a,b,\cdots$ run the directions of the two-sphere.
In the static and spherically symmetric spacetime Eq. \eqref{sss},
only the $t$- and $r$-components of the vector field 
can have nonzero values  
\begin{eqnarray}
\label{vector_sss}
A_\mu dx^\mu
=A_t (r)dt+ A_r(r) dr,
\end{eqnarray}
with $X=-e^{-\nu}A_t^2+e^{-\lambda} A_r^2$.

We focus on the radial perturbations
where the vector field perturbation is explicitly given by 
\begin{eqnarray}
\delta A_\mu dx^\mu
=\delta A_t (t,r)dt+ \delta A_r(t,r) dr.
\end{eqnarray}
We note that
for the nonradial perturbations with the multipole indices $\ell$,
because of the centrifugal term proportional to $\ell (\ell+1)$,
the effective potential is always enhanced,
and hence to discuss the stability of the GR solutions
it is sufficient to discuss the radial perturbations.
In the static GR background \eqref{sss} with $A_t=A_r=0$
and $u^\mu\propto \delta^\mu_t$,
at the level of the linearized perturbations
we obtain
\begin{eqnarray}
\label{linear1}
\partial_r (\delta F_{r t})
+
\left(
\frac{2}{r}
-\frac{\nu'}{2}
-\frac{\lambda'}{2}
\right) 
\delta F_{r t}
-
{\cal D}_t
e^\lambda
\delta A_t
&=&
0,
\\
\label{linear2}
\partial_t (\delta F_{r t})
-{\cal D}_r e^\nu
\delta A_r
&=&
0,
\end{eqnarray} 
where 
the prime denotes the derivative with respect to $r$,
$\delta F_{rt}=\partial_r \delta A_t-\partial_t \delta A_ r$,
and we have defined
\begin{eqnarray}
{\cal D}_t
&:=&
m^2
+2\kappa^2
\left[
B_0 \rho 
+C_1
 \left(
\rho-3p
 \right)
\right],
\\
{\cal D}_r
&:=&
  m^2
+2\kappa^2 
\left[
-B_0 p
 +C_1 (\rho-3p)
\right].
\end{eqnarray}
Combining Eqs. \eqref{linear1} and \eqref{linear2},
we obtain the master equation for the vector field perturbation
about the $A_t=A_r=0$ background
\begin{widetext}
\begin{eqnarray}
\partial_r
\left[
\frac{e^{-\lambda}}
       {{\cal D}_t}
\left(
\partial_r
(\delta F_{rt})
+
\left(
\frac{2}{r}
-\frac{\nu'}{2}
-\frac{\lambda'}{2}
\right)
\delta F_{rt}
\right)
\right]
-\frac{e^{-\nu}}{{\cal D}_r}
  \partial_t^2 (\delta F_{rt})
-\delta F_{rt}
=0.
\end{eqnarray}
\end{widetext}
In the case of the pure vector conformal coupling $B_0=0$ and $C_1\neq 0$,
${\cal D}_r={\cal D}_r=m^2+2\kappa^2 C_1 (\rho-3p)$
and for $\rho-3p>0$
the tachyonic instability would take place for $C_1<0$
even for $m\geq 0$.
The tachyonic instability 
suggests that 
the GR solution would spontaneously evolve to the nontrivial solution
with the nonzero value of the vector field,
which would be energetically more stable.
In Sec. \ref{sec7},
we will construct the nontrivial solutions
for the massless case $m=0$
as the possible consequence of the tachyonic instability.

On the other hand, 
in the case of the pure vector disformal coupling $B_0\neq 0$ and $C_1=0$,
${\cal D}_t=m^2+2\kappa^2 B_0\rho$
and 
${\cal D}_r=m^2-2\kappa^2 B_0 p$,
and hence 
especially for the massless case $m=0$,
we have ${\cal D}_t {\cal D}_r\propto - B_0^2 \rho p <0$,
irrespective of the sign of $B_0$.
This suggests
that the GR solution 
would suffer from the ghost or gradient instability,
rather than the tachyonic instability.
This instability
would be related to the breakdown of the hyperbolicity on the GR stellar backgrounds,
rather than the mere energetic instability.
Thus, 
the pure disformal coupling would not cause 
spontaneous vectorization
as in the way analogous to spontaneous scalarization.
The result is rather different from the expectation in Ref. \cite{,Ramazanoglu:2019jrr}
that the disformal coupling to the matter would also cause
the tachyonic instability on the GR stellar background.
Nevertheless, 
in Sec. \ref{sec6},
we will explicitly construct the solutions of the relativistic stars
with the nontrivial profile of the vector field,
as they may be formed 
via the initial conditions different from the GR solutions with $A_\mu =0$. 
We will explicitly confirm 
that in the mass-radius diagram
 the branches of the nontrivial solutions 
are disconnected to that of the GR solutions.

\section{Static and spherically symmetric spacetime}
\label{sec3}

In this section, 
we focus on the static and spherically symmetric spacetime \eqref{sss} and \eqref{vector_sss},
and derive the equations to analyze the hydrostatic structure of the relativistic stars,
i.e., the modified Tolman-Oppenheimer-Volkoff equations.

\subsection{The modified Tolman-Oppenheimer-Volkoff equations}

We assume that 
the nonzero components in the Jordan frame \eqref{disformal}
are given by the perfect fluid form
\begin{eqnarray}
{\tilde T}^t{}_t
=-{\tilde \rho},
\qquad
{\tilde T}^r{}_r
={\tilde p}_r,
\qquad
{\tilde T}^a{}_b
={\tilde p}_t \delta^a{}_b.
\end{eqnarray}
Correspondingly,
the nonzero components of the energy-momentum tensor of matter in the Einstein frame are given by
Eq. \eqref{em_sss}.

The $r$-component of the vector field equation \eqref{veq}
is given by 
\begin{widetext}
\begin{eqnarray}
A_r
\left\{
-m^2
+
\frac{2\kappa^2 C^2}{\sqrt{1+BX}}
\left[
B{\tilde p}_r 
+B_X
\left(
e^{-\nu}
{\tilde\rho} A_t^2
+e^{-\lambda}
{\tilde p}_r A_r^2
\right)
+\frac{C_X}{C}(1+BX)
\left(
-{\tilde \rho}+{\tilde p}_r+2{\tilde p}_t
\right)
\right]
\right\}
=0,
\end{eqnarray}
\end{widetext}
which allows the two branches,
$A_r=0$ or the combination inside the round bracket vanishes.
Here, we choose the former branch $A_r=0$,
and then 
the Jordan frame metric Eq. \eqref{jordan_sss} reduces to  
the diagonal form
{
and 
the energy-momentum tensor of matter in the Einstein frame
also has the perfect fluid form
whose components are given by Eq. \eqref{perfect_eins}.

The $t$-component of the vector field equation of motion \eqref{veq} is given by 
\begin{widetext}
\begin{eqnarray}
\label{veq_t}
&& 
A_t''
+\left(\frac{2}{r}-\frac{\nu'}{2}-\frac{\lambda'}{2} \right)
A_t'
-e^{\lambda}A_t
\left\{
 m^2
+
\frac{2\kappa^2 C^2}{\sqrt{1+BX}}
\left[
{\tilde\rho} 
(B+B_X X)
-\frac{C_X}{C}(1+BX)
\left(
-{\tilde \rho}+{\tilde p}_r+2{\tilde p}_t
\right)
\right]
\right\}
=0.
\end{eqnarray}
On the other hand,  
the $(t,t)$-, $(r,r)$-, and $(a,b)$- components of the 
gravitational equations of motion \eqref{eeq}
are respectively given by 
\begin{eqnarray}
\label{eq1}
&&
\frac{1}{r^2}
\left[
1- e^{-\lambda }(1-r\lambda') 
\right]
=
 \frac{1}{4}e^{-\lambda-\nu} A_t'^2
-\frac{m^2}{4}X
+\frac{\kappa^2 C^2}{\sqrt{1+BX}}
\left\{
 \tilde\rho
\left(
1+B_X X^2
\right)
-\frac{C_X X}{C}
(1+BX)
\left(
-{\tilde \rho}
+{\tilde p}_r
+2{\tilde p}_t
\right) 
\right\},
\\
\label{eq2}
&&
\frac{1}{r^2}
\left[
1- e^{-\lambda}(1+r\nu' )
\right]
=
\frac{1}{4}e^{-\lambda-\nu} A_t'^2
+\frac{m^2}{4}X
-\kappa^2 C^2
\sqrt{1+BX} {\tilde p}_r,
\\
\label{eq3}
&&
\frac{e^{-\lambda}}{2}
\left[
\nu''
+
\left(\frac{1}{r}+\frac{\nu'}{2}\right)
(\nu'-\lambda')
\right]
=
\frac{1}{4}e^{-\lambda-\nu} A_t'^2
-\frac{m^2}{4}X
+\kappa^2 C^2 \sqrt{1+BX}
{\tilde p}_t.
\end{eqnarray}
Finally, 
the nontrivial $r$-component of Eq. ~\eqref{ceq} is given by 
\begin{eqnarray}
\label{pressure_eq}
&&
{\tilde p}_r'
+\frac{1}{2(1+BX)}
\left[
{\tilde p}_r+ {\tilde \rho}
+B_X X^2 
  (-{\tilde p}_r+{\tilde \rho})
-\frac{C_X X}{C}
 (1+BX) 
  \left(
  5{\tilde p}_r
+2{\tilde p}_t
-{\tilde \rho}
 \right)
\right]\nu'
\nonumber\\
&-&
e^{-\nu/2}
\frac{\sqrt{-X}}
      {1+BX}
\left[
({\tilde p}_r+{\tilde\rho})
(B+B_X X)
+(1+BX) (3{\tilde p}_r-2{\tilde p}_t+{\tilde\rho})
\frac{C_X}{C}
\right]
A_t'
+
\frac{2}{r}
\left(
 {\tilde p}_r
- {\tilde p}_t
\right)
=0.
\end{eqnarray}
\end{widetext}

Eliminating $\lambda$ with the mass function $\mu$ defined by 
\begin{eqnarray}
\label{mass_function}
e^{-\lambda}
=1-\frac{2\mu(r)}{r},
\end{eqnarray}
Eqs. \eqref{eq1} and \eqref{eq2} can be rewritten as
\begin{eqnarray}
\label{dep1}
\mu'
&=&
\frac{r^2}{8}
\left(
1-\frac{2\mu}{r}
\right)
e^{-\nu} A_t'^2
-
\frac{m^2r^2}{8}X
\nonumber\\
&+&
\frac{\kappa^2 C^2 r^2}{2\sqrt{1+BX}}
\left\{
 \tilde\rho
\left(
1+B_X X^2
\right)
\right.
\nonumber\\
&&
\left.
-\frac{C_X X}{C}
(1+BX)
\left(
-{\tilde \rho}
+{\tilde p}_r
+2{\tilde p}_t
\right) 
\right\},
\\
\label{dep2}
\nu'
&=&
\frac{r^2}{r-2\mu}
\left[
 \frac{2\mu}{r^3}
-\frac{m^2}{4}X
+\kappa^2 C^2
\sqrt{1+BX}{\tilde p}_r
\right]
\nonumber\\
&-&
\frac{r}{4}e^{-\nu}A_t'^2.
\end{eqnarray}
Eq. \eqref{pressure_eq} is rewritten as 
\begin{widetext}
\begin{eqnarray}
\label{dep3}
{\tilde p}_r'
&=&
-\frac{
{\tilde p}_r+ {\tilde \rho}
+B_X X^2 
  (-{\tilde p}_r+{\tilde \rho})
-\frac{C_X X}{C}
 (1+BX) 
  \left(
  5{\tilde p}_r
+2{\tilde p}_t
-{\tilde \rho}
 \right)}
{2(1+BX)}
\left\{
\left[
 \frac{2\mu}{r^3}
-\frac{m^2}{4}X
+\kappa^2 C^2
\sqrt{1+BX}{\tilde p}_r
\right]
-\frac{r}{4}e^{-\nu}A_t'^2
\right\}
\nonumber\\
&+&
e^{-\nu/2}
\frac{\sqrt{-X}}
      {1+BX}
\left[
({\tilde p}_r+{\tilde\rho})
(B+B_X X)
+(1+BX) (3{\tilde p}_r-2{\tilde p}_t+{\tilde\rho})
\frac{C_X}{C}
\right]
A_t'
-
\frac{2}{r}
\left(
 {\tilde p}_r
- {\tilde p}_t
\right).
\end{eqnarray}
Similarly, eliminating $\lambda$,
Eq. \eqref{veq_t} reduces to 
\begin{eqnarray}
\label{dep4}
&& 
A_t''
+\left\{
\frac{2}{r}
-\frac{1}{2} \nu'
+\frac{1}{r(r-2\mu)}
\left(
\mu -r\mu'
\right)
\right\}
A_t'
\nonumber\\
&-&
\frac{r}{r-2\mu}
A_t
\left\{
 m^2
+
\frac{2\kappa^2 C^2}{\sqrt{1+BX}}
\left[
{\tilde\rho} 
(B+B_X X)
-\frac{C_X}{C}(1+BX)
\left(
-{\tilde \rho}+{\tilde p}_r+2{\tilde p}_t
\right)
\right]
\right\}
=0,
\end{eqnarray}
\end{widetext}
with the substitution of Eqs. \eqref{dep1} and \eqref{dep2}.

We focus on matter with the isotropic pressure 
\begin{eqnarray}
{\tilde p}:={\tilde p}_r={\tilde p}_t,
\end{eqnarray}
and then 
all the unknown are the five variables $\nu$, $\lambda$, ${\tilde \rho}$, ${\tilde p}$, 
and $A_t$,
while we have the four independent equations
Eqs. \eqref{dep1}, \eqref{dep2}, \eqref{dep3}, and \eqref{dep4}.
The system of the equations is closed 
once the equation of state
\begin{eqnarray}
\label{equation of state}
{\tilde \rho}={\tilde \rho}({\tilde p}),
\end{eqnarray}
is specified.
We note that Eq. \eqref{eq3}  is not independent of the other equations.

\subsection{The interior solution}

The interior solution is obtained 
by integrating Eqs. \eqref{dep1}, \eqref{dep2}, \eqref{dep3}, and \eqref{dep4} 
with a given equation of state \eqref{equation of state}
from the center of the star $r=0$ to the surface of the star $r=\mR$
whose position is determined by the condition 
(See Eq. \eqref{perfect_eins})
\begin{eqnarray}
\label{surface}
p(\mR)
={\tilde p}(\mR)=0.
\end{eqnarray}  
The interior solution in the vicinity of the center is given by 
\begin{eqnarray}
\label{center}
\mu_{(i)}
&=&
\mu_3 r^3+{\cal O} (r^5),
\nonumber\\
\nu_{(i)}
&=&\nu_0 + \nu_2 r^2+{\cal O} (r^4),
\nonumber \\
A_{t(i)}
&=& A_C +a_2 r^2+{\cal O} (r^4),
\nonumber\\
{\tilde p}
&=&
{\tilde p}_0 +{\tilde p}_2 r^2 +{\cal O}(r^4),
\end{eqnarray}
where the coefficients
$\mu_3$, $\nu_2$, $a_2$, and ${\tilde p}_2$
are given in Eq. \eqref{central},
with $X_0:= A_C^2 e^{\nu_0/2}$, $B=B(X_0)$, $B_X=B_X(X_0)$, $C=C(X_0)$, and $C_X=C_X(X_0)$
in these relations.
Without loss of generality, we set $\nu_0=0$,
so that 
the time coordinate $t$ corresponds to the proper time at the center of the star. 
From Eq. \eqref{central}, 
the case of $m=0$, $C(X)=1$, and $B(X)=0$
recovers the stellar solution in GR
\begin{eqnarray}
&&
\mu_3
=
\frac{\kappa^2}{6}
{\tilde \rho}_0,
\qquad
\nu_2
=
\frac{\kappa^2}{6}
\left(
{\tilde \rho}_0
+3{\tilde p}_0
\right),
\nonumber\\
&&
{\tilde p}_2
=
-\frac{\nu_2}{2}
({\tilde\rho}_0+{\tilde p}_0).
\end{eqnarray}
From Eq.
\eqref{jordan_sss} with $A_r=0$,
we find the relation of the mass functions and radial coordinates
of the Jordan and Einstein frames,
where
${\tilde g}_{rr}
=1/(1-2{\tilde \mu}/{\tilde r})$,
and 
${\tilde g}_{ab}
={\tilde r}^2\gamma_{ab}$
respectively given by 
\begin{eqnarray}
{\tilde r}
&=&
\sqrt{C(X)}r,
\\
{\tilde \mu}
&=&
\frac{\sqrt{C (X)}r }{2}
\left[
1
-
\left(
1+\frac{r C_X X}{2C}
\right)^2
\left(
1-\frac{2\mu}{r}
\right)
\right].
\end{eqnarray}
Thus, 
the radius of the star in both the frames is related by 
\begin{eqnarray}
\label{radius}
{\tilde \mR}
&=&
\sqrt{C(X)|_{r=\mR}}\mR.
\end{eqnarray}
We assume that 
as $r\to \infty$, 
$X\to 0$, $C(X)\to 1$, and $C_X$ is regular, 
and then the masses in both the frames coincide
\begin{eqnarray}
\label{mass}
{\tilde M}
&=&
M.
\end{eqnarray}

\subsection{The external solution and the matching at the surface of the star}

From now on, we set $m=0$.
The external solution is then given by
the Reissner-Nordstr\"om solution
\begin{eqnarray}
&&
e^{-\lambda_{(o)}}
=1-\frac{2GM}{r}
+\frac{G^2Q^2}{4r^2},
\nonumber
\\
&&
e^{\nu_{(o)}}
=
q_\infty^2
\left(
1-\frac{2G M}{r}
+\frac{G^2 Q^2}{4r^2}
\right),
\nonumber\\
&&
A_{(o) t}
=
q_\infty
\left(P+\frac{GQ}{r}\right),
\end{eqnarray}
where $M$,~$P$,~$Q$, and $q_\infty$ are integration constants.
The external solution is matched
to the internal one at the surface of the star $r=\mR$ determined by Eq. \eqref{surface}.
Imposing the continuity of  $\nu$, $\nu'$, $A_t$ and $A_t'$ at $r=\mR$, 
\begin{eqnarray}
&&
\nu_{(o)}(\mR)=\nu_{(i)} (\mR),
\qquad 
\nu_{(o)}'(\mR)=\nu_{(i)}' (\mR),
\nonumber
\\
&&
A_{t(o)}(\mR)=A_{t(i)} (\mR),
\qquad 
A_{t(o)}'(\mR)=A_{t(i)}' (\mR),
\end{eqnarray}
we obtain
\begin{eqnarray}
\label{charges}
&&
P= \frac{A_{t(i)} (\mR)+\mR A_{t(i)}'(\mR)}
          {\sqrt{\frac{1}{4}\mR^2 A_{t(i)}'(\mR)^2+ e^{\nu_{(i)} (\mR)} \left(1+\mR \nu_{(i)}' (\mR)\right)}},
\nonumber\\
&&M= 
\frac{1}{G}
\frac{\mR^2 \left(\mR A_{t(i)}'(\mR)^2+ 2e^{\nu_{(i)} (\mR)}  \nu_{(i)}' (\mR)\right)}
          {\mR^2 A_{t(i)}'(\mR)^2+ 4e^{\nu_{(i)} (\mR)} \left(1+\mR \nu_{(i)}' (\mR)\right)},
\nonumber\\
&&
Q=
-\frac{1}{G}
 \frac{\mR^2 A_{t(i)}'(\mR)}
        {\sqrt{\frac{1}{4}\mR^2 A_{t(i)}'(\mR)^2+ e^{\nu_{(i)} (\mR)} 
\left(1+\mR \nu_{(i)}' (\mR)\right)}},
\nonumber\\
&&
q_\infty
=
\sqrt{
\frac{1}{4} \mR^2 A_{t(i)}'(\mR)^2
+e^{\nu_{(i)} (\mR)} 
   \left(1+\mR \nu_{(i)}' (\mR)\right)}.
\end{eqnarray}
$M$ and $Q$ correspond to the mass and  vector field charge of the star.
Since the gauge invariance is broken because of the vector field coupling to matter,
$P$ is also a physical quantity.
The proper time measured at the spatial infinity
is given by $dt_\infty =q_\infty dt$.
The redshift factor for photons radially
traveling from the center of the star to the spatial infinity
is given by 
\begin{eqnarray}
z_0 :=\frac{\omega_0}{\omega_\infty}-1
 =q_\infty-1,
\end{eqnarray}
where $\omega_0$ and $\omega_\infty$
are the frequencies measured at the center of the star and 
at the spatial infinity, respectively.
Similarly,
the redshift between the surface of the star and the spatial infinity
is given by 
\begin{eqnarray}
z_R
&:&=\frac{\omega_R}{\omega_\infty}-1
 =q_\infty e^{-\nu_{(i)} (\mR)/2}
-1
\nonumber\\
&=&
\sqrt{
 1+\mR \nu_{(i)}' (\mR)
+\frac{\mR^2}{4} A_{t(i)}'(\mR)^2
e^{-\nu_{(i)} (\mR)}}
-1,
\end{eqnarray}
where $\omega_R$ represents the frequency measured at the surface of the star.

\subsection{The polytrope equation of state}

For the numerical analyses in Secs. \ref{sec5}, \ref{sec6}, and \ref{sec7},
we employ the polytrope equation of state
\begin{eqnarray}
\label{poly1}
{\tilde p}={\cal K} \rho_0^\Gamma,
\end{eqnarray}
where $\rho_0=m_b n$ is the rest mass density
with $m_b$ and $n$ being the baryonic mass and the number density,
respectively, 
and 
$\Gamma$ and ${\cal K}$ are constants. 
The mean baryonic mass is identified with the neutron mass $m_b=1.6749\times 10^{-24}{\rm g}$.

The total energy density is given by 
${\tilde \rho}=\rho_0 (1+\epsilon)$,
where $\epsilon$ is the dimensionless internal energy per unit mass.
For adiabatic nuclear matter,  
the first law of thermodynamics
reduces to $d{\tilde \rho}= ({\tilde\rho}+{\tilde p}) dn/n$.
With 
$d{\tilde \rho}= m_b(1+\epsilon )dn+m_b n d\epsilon$
and 
$({\tilde\rho}+{\tilde p}) dn/n=m_b(1+n)dn+{\tilde p}dn/n $,
the first law leads to ${\tilde p}=m_b n^2 \partial \epsilon /\partial n$,
which with Eq. \eqref{poly1} can be integrated as
$\epsilon ={\cal K} \rho_0^{\Gamma-1}/(\Gamma-1)$.
Thus, the relation between the total energy density and pressure in the Jordan frame 
is given by  
\begin{eqnarray}
\label{kspace}
{\tilde \rho}
=\rho_0 
+\frac{{\cal K} \rho_0^{\Gamma}}
       {\Gamma-1}
=\left(\frac{\tilde p}{\cal K}\right)^{\frac{1}{\Gamma}}
+\frac{\tilde p}{\Gamma-1}.
\end{eqnarray}
Defining the dimensionless number density by 
$\chi= \rho_0/  \bar{\rho}_0  = n/{\bar n}$,
with ${\bar\rho}_0=m_b {\bar n}$ and ${\bar n}=0.1 ({\rm fm})^{-3}$ being the mean number density of the baryonic particles,
the total energy density and pressure can be written as 
\begin{eqnarray}
{\tilde \rho}
&=&
{\bar\rho}_0
\left(
\chi
+\frac{K}{\Gamma-1}
\chi^\Gamma
\right),
\qquad
{\tilde p}
=
K {\bar \rho}_0
\chi^\Gamma,
\end{eqnarray}
where $K:={\cal K}/ ({\bar\rho}_0)^{1-\Gamma}$.
Eq. \eqref{kspace} can be rewritten in terms of $K$
\begin{eqnarray}
\label{pressure}
  {\tilde \rho}
={\bar \rho}_0
 \left(
  \frac{\tilde p} {K{\bar\rho}_0}
 \right)^{\frac{1}{\Gamma}}
+\frac{{\tilde p}}{\Gamma-1}.
\end{eqnarray}
For our numerical analyses, 
we will choose the parameters
\begin{eqnarray}
K=0.013, 
\qquad 
\Gamma=2.34.
\end{eqnarray}
We also parametrize the pressure at the center of the star ($r=0$) as
\begin{eqnarray}
{\tilde p}_C:=
{\tilde p}(0)
&=&
 j \times 10^{14}  
[{\rm g}\cdot {\rm cm}^{-3}]\times c^2
\nonumber\\
&=&
9.0\times j \times 10^{35}  
[{\rm dyne}\cdot {\rm cm}^{-2}],
\end{eqnarray}
where $j$ represents a constant parameter.
Correspondingly, 
from Eq. \eqref{pressure}
then the energy density at the center of the star is given by 
\begin{eqnarray}
\label{pressure_center}
  {\tilde \rho}_C
:={\tilde\rho} (0)
=
{\bar \rho}_0
 \left(
  \frac{{\tilde p}_C} 
           {K{\bar\rho}_0}
 \right)^{\frac{1}{\Gamma}}
+\frac{{\tilde p}_C}
         {\Gamma-1}.
\end{eqnarray}

\section{The test vector field solutions in the weak gravity regimes}
\label{sec4}

Before going to the numerical studies with the polytrope
equation of state \eqref{poly1} and \eqref{pressure}
in the next sections,
we estimate the critical vector field couplings 
that enhance the amplitude and charge 
of the test vector field
on top of the constant density star in GR.

For any regular coupling \eqref{regular_coupling}, 
the GR stellar solution with $A_t=A_r=0$ 
exists in the vector-tensor theory \eqref{action} with Eq. \eqref{disformal}.  
Since there is no distinction between the Jordan and Einstein frames,
$\rho={\tilde \rho}$,
and 
$p={\tilde p}$.
We consider an incompressible fluid, ${\tilde \rho}=\rho_0={\rm const}$,
where the interior metric and pressure ($r<\mR$) are given by 
\begin{eqnarray}
\label{gr_exa}
e^{\lambda(r)}
&=&\left(1-\frac{2GM_0 r^2}{\mR^3}\right)^{-1},
\nonumber\\
e^{\nu(r)}
&=&
\left[
\frac{3}{2}
 \left(1-\frac{2GM_0}{\mR}\right)^{1/2}
-\frac{1}{2}
\left(1-\frac{2GM_0 r^2}{ \mR^3}\right)^{1/2}
\right]^2,
\nonumber\\
{\tilde p}(r)&=&\rho_0 
\frac{\left(1-\frac{2GM_0 r^2}{ \mR^3}\right)^{1/2}
      -\left(1-\frac{2GM_0}{ \mR}\right)^{1/2}}
      {3\left(1-\frac{2GM_0}{\mR}\right)^{1/2}
        -\left(1-\frac{2GM_0 r^2}{\mR^3}\right)^{1/2}},
\end{eqnarray}
where at the surface of the star $r=\mR$, ${\tilde p}(\mR)=0$,
and $M_0$ and ${\cal C}$ are the total mass and compactness
\begin{equation}
M_0=\frac{4\pi \mR^3}{3}\rho_C,
\qquad
{\cal C}=\frac{GM_0}{\mR}.
\end{equation}
The exterior solution ($r>\mR$) is given by
the Schwarzschild metric 
\begin{eqnarray}
e^{\lambda(r)}
&=&
e^{-\nu(r)}
=\left(1-\frac{2GM_0}{r}\right)^{-1}.
\end{eqnarray}
On top of the constant density star,
the test vector field obeys the equation 
\begin{eqnarray}
&&
A_t''
+\left(
\frac{2}{r}
-\frac{\nu'}{2}
-\frac{\lambda'}{2} 
\right)
A_t'
\nonumber\\
&+&
e^{\lambda}
2\kappa^2
\left[
 B_0 {\tilde\rho} 
+C_1
\left(
{\tilde \rho}-3{\tilde p}
\right)
\right]
A_t
=0.
\end{eqnarray}
Thus, in the weak gravity regime ${\cal C}\ll 1$,
the approximated interior solution ($0<r<\mR$) is given by 
\begin{eqnarray}
\label{int}
A_{t} (0<r<\mR)
\approx
A_C
\frac{\sin(\sqrt{-2 (B_0+C_1)\rho_C} \kappa r )}
       {\sqrt{-2 (B_0+C_1)\rho_C} \kappa r },
\end{eqnarray}
where we have chosen the regularity boundary conditions
at the center of the star $r=0$,
for which the solution decreases.
We assume that 
\begin{eqnarray}
B_0+C_1<0.
\end{eqnarray}
We note that for $B_0+C_1>0$,
the interior solution regular at the center of the star
grows as 
$\sinh(\sqrt{2 (B_0+C_1)\rho_C} \kappa r )/r$,
which is unphysical and excludes this possibility in the rest.

The exterior solution ($r>\mR$) is given by 
\begin{eqnarray}
\label{ext}
A_t(r>\mR)
=P+ \frac{GQ}{r}.
\end{eqnarray} 
The matching of Eqs. \eqref{int} and \eqref{ext} 
at the surface of the star $r=\mR$ 
provides
\begin{eqnarray}
&&
\frac{A_C}{P}
\simeq 
\frac{1}
      {\cos(\sqrt{-2 (B_0+C_1)\rho_C} \kappa \mR )},
\nonumber\\
&&
\frac{GQ}{\mR}\simeq 
P
\left[
-1
+\frac{\tan(\sqrt{-2 (B_0+C_1)\rho_C} \kappa \mR )}
         {\sqrt{-2( B_0+C_1)\rho_C} \kappa \mR }
\right].
\end{eqnarray}
As $\sqrt{-2 (B_0+C_1)\rho_C} \kappa \mR =\sqrt{-12(B_0+C_1) {\cal C}}$,
$Q$ and $A_C$ are enhanced
for the case
\begin{eqnarray}
\label{upper}
-\left(B_0+C_1\right)
\approx 
\frac{\pi^2}{48 {\cal C}} (1+2n)^2,
\end{eqnarray}
where $n=0,1,2,\cdots$ denotes the number  of nodes of the vector field inside the star.
The estimation above is analogous 
to the case of spontaneous scalarization \cite{Harada:1997mr}.

Here, we focus on the fundamental $0$ node solutions $n=0$.
Extrapolating to the relatively high compact star ${\cal C}={\cal O} (0.1)$, 
the value of the coupling for which the vector field is amplified is given by 
$-(B_0+C_1) ={\cal O} (1.0)$. 
If the above static solution becomes the weak field approximation
for the fully vectorized relativistic stars,
it implies that  
the full vectorized solution would exist for  $(-B_0)$ or $(-C_1)$ 
to be of ${\cal O} (1.0)$.

\section{The case of the pure disformal coupling}
\label{sec5}

In this section,
we consider the simplest model with the pure disformal coupling given by  
\begin{eqnarray}
\label{pure_disformal}
C(X)=1, 
\qquad 
B(X)=B_0,
\end{eqnarray}
where $B_0$ is the constant.
From Eqs.  \eqref{radius} and \eqref{mass},
we obtain 
\begin{eqnarray}
{\tilde \mR}=\mR,
\qquad 
{\tilde M}=M.
\end{eqnarray}
As argued in Sec. \ref{sec22},
the pure disformal coupling
induces the ghost or gradient instability of the GR solutions,
rather than the tachyonic instability.
Thus,
the hyperbolicity of the perturbations is broken on the GR stellar background 
and hence
spontaneous vectorization analogous to spontaneous scalarization
would not take place.
Nevertheless,
we will study the nontrivial solutions satisfying the boundary condition $P=0$,
as they may be formed from another choice of the initial conditions.
As we will see below,
the properties of the solutions are similar to 
those obtained in the generalized Proca theories \cite{Kase:2020yhw}.

\subsection{The $0$ node solutions}

First, we focus on the $0$ node solutions,
where the temporal component of the vector field $A_t$ never crosses $0$ both the interior and exterior of the star. 
In Fig. \ref{0node}, 
the typical behavior for the $0$ node solutions is shown for $B_0=-0.5$.
In both the panels,
the solid and dashed curves correspond to the cases of $j=1.0$ and $j=2.0$, respectively.
In the top panel,
the red and blue curves
represent
the energy density ${\tilde \rho}c^2$ and pressure ${\tilde p}$ inside the star,
respectively,
and the horizontal and vertical axes are shown 
in the units of ${\rm cm}$ and ${\rm dyne}\cdot {\rm cm}^{-2}$,
respectively.
The bottom panel shows
the temporal component of the vector field $A_t$
as the function of $r/{\cal R}$.
In the bottom panel,
the red points correspond to the surface of the star.
\begin{figure}[h]
\unitlength=1.1mm
\begin{center}
  \includegraphics[height=4.5cm,angle=0]{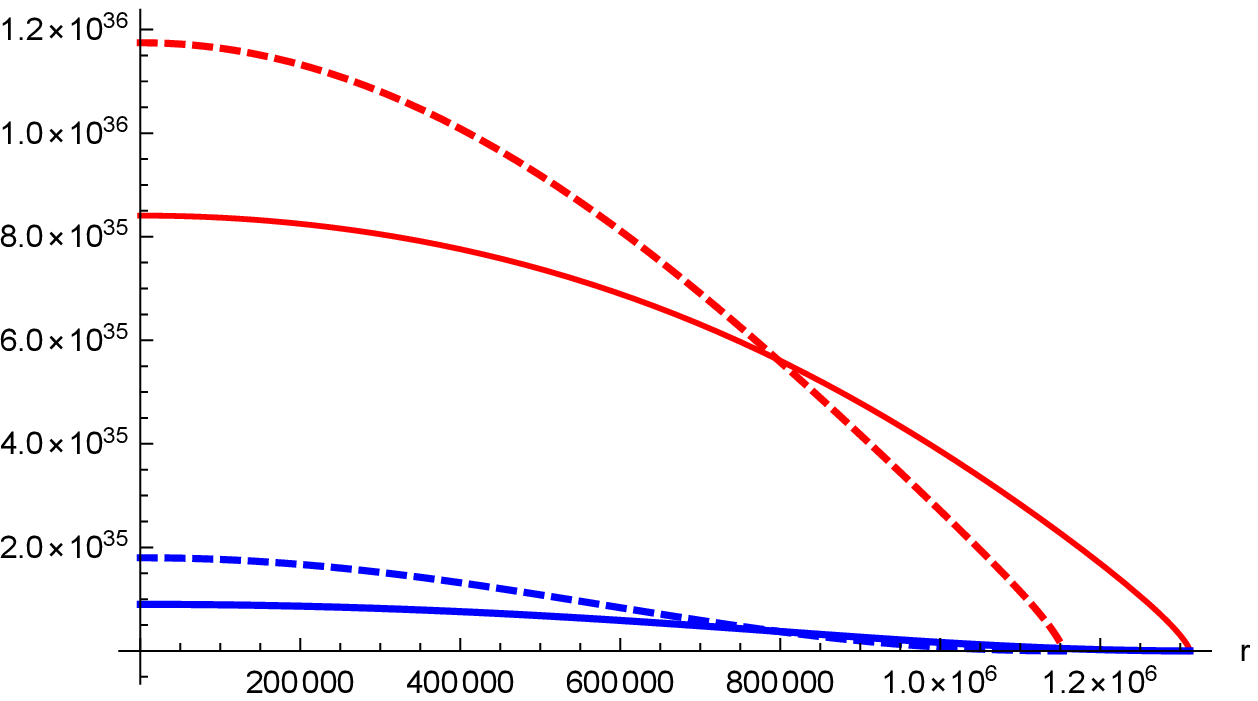}  
  \includegraphics[height=4.5cm,angle=0]{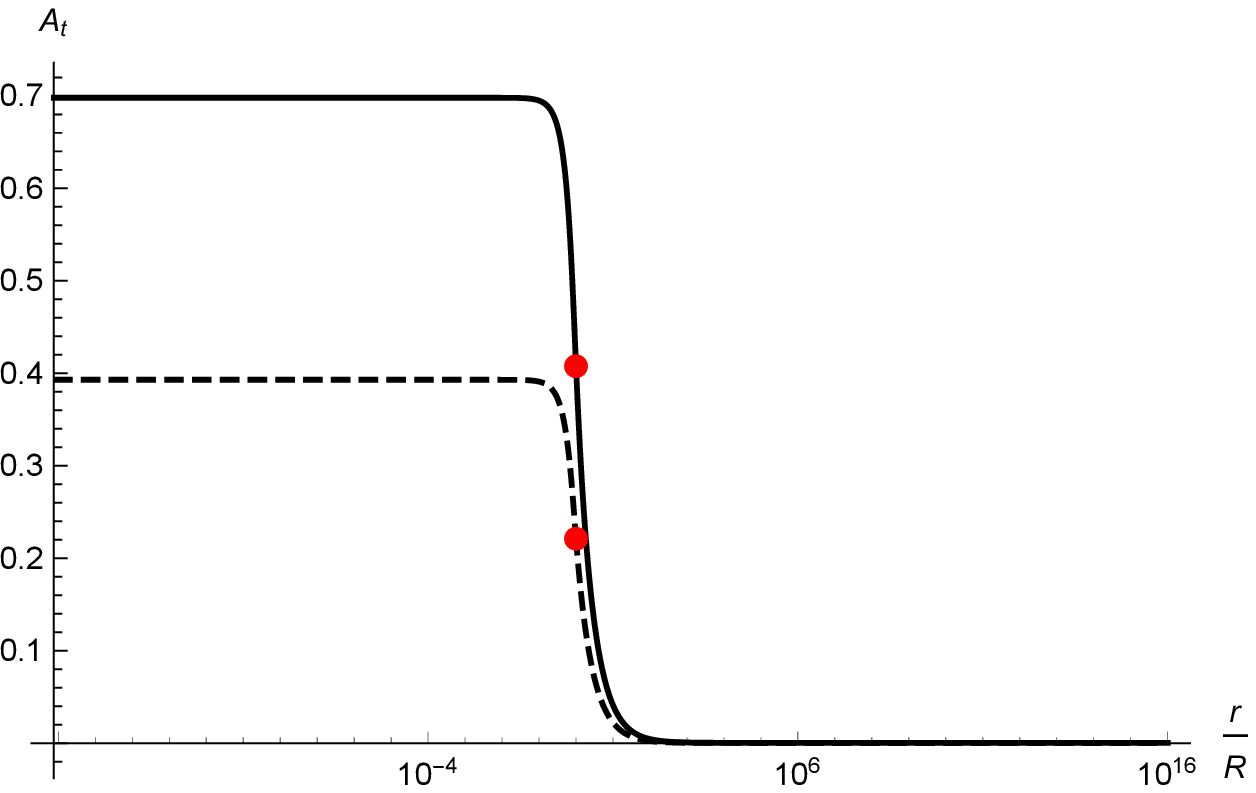}
\caption{
The typical behavior for the $0$ node solutions is shown for $B_0=-0.5$.
In both the panels,
the solid and dashed curves correspond to the cases of $j=1.0$ and $j=2.0$, respectively.
In the top panel,
the red and blue curves
represent
the energy density ${\tilde \rho}c^2$ and pressure ${\tilde p}$ inside the star,
respectively,
and the horizontal and vertical axes are shown 
in the units of ${\rm cm}$ and ${\rm dyne}\cdot {\rm cm}^{-2}$,
respectively.
The bottom panel shows
the temporal component of the vector field $A_t$
as the function of $r/{\cal R}$.
In the bottom panel,
the red points correspond to the surface of the star.
}
  \label{0node}
\end{center}
\end{figure} 

In Fig.~\ref{figroot0node},
following Eq. \eqref{charges},
$P$ is shown as the function of $A_C$
for $B_0=-0.5$. 
The red, blue-dashed, green-dotted, and black-dotdashed curves
correspond to
the cases of $j=0.5, 1.0, 2.0, 3.0$,
respectively.
\begin{figure}[h]
\unitlength=1.1mm
\begin{center}
  \includegraphics[height=5.1cm,angle=0]{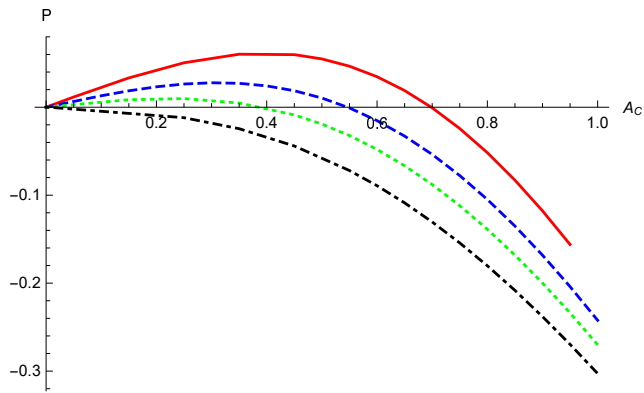}
\caption{
$P$ is shown as the function of $A_C$ for $B_0=0.5$. 
The red, blue-dashed, green-dotted, and black-dotdashed curves 
correspond to
the cases of $j=0.5, 1.0, 2.0, 3.0$, respectively.
}
  \label{figroot0node}
\end{center}
\end{figure} 
All the curves take $P=0$ at $A_C=0$,
which corresponds to the GR solutions.
The curves for $j=0.5, 1.0, 2.0$
take the positive values for the smaller $A_C$,
and then cross the axis of $P=0$ once,
where the $0$ node solutions exist.
On the other hand,
the curve for $j=3.0$
never crosses the axis of $P=0$ except at $A_C=0$
and hence there exists only the GR solutions.
As $j$, i.e., ${\tilde p}_C$, increases, 
the root of $P=0$ 
becomes smaller
and
finally reaches the origin $A_C=0$.
Thus, above the critical value of $j$, 
the $0$ node solutions converge to the GR solutions with $A_C=0$.

In Fig.~\ref{fig0node},
the mass-radius diagram is shown for the $0$ node solutions.
In the top panel, 
the red, blue-dashed, and green-dotted curves 
correspond to
the cases of $B_0=-0.3,-0.5,-0.7$,
respectively,
while the black curve corresponds to the GR solutions.
The bottom panel focuses on the case of $B_0=-0.5$.
In the bottom panel, 
the black-dashed curve represents
the range of the branch of the GR solutions
for which the branch of the $0$ node solutions coexists.
The radius and mass are measured 
in the units of ${\rm km}$ and
the Solar mass $M_\odot=1.988\times 10^{33} {\rm g}$,
respectively.
\begin{figure}[h]
\unitlength=1.1mm
\begin{center}
  \includegraphics[height=5.1cm,angle=0]{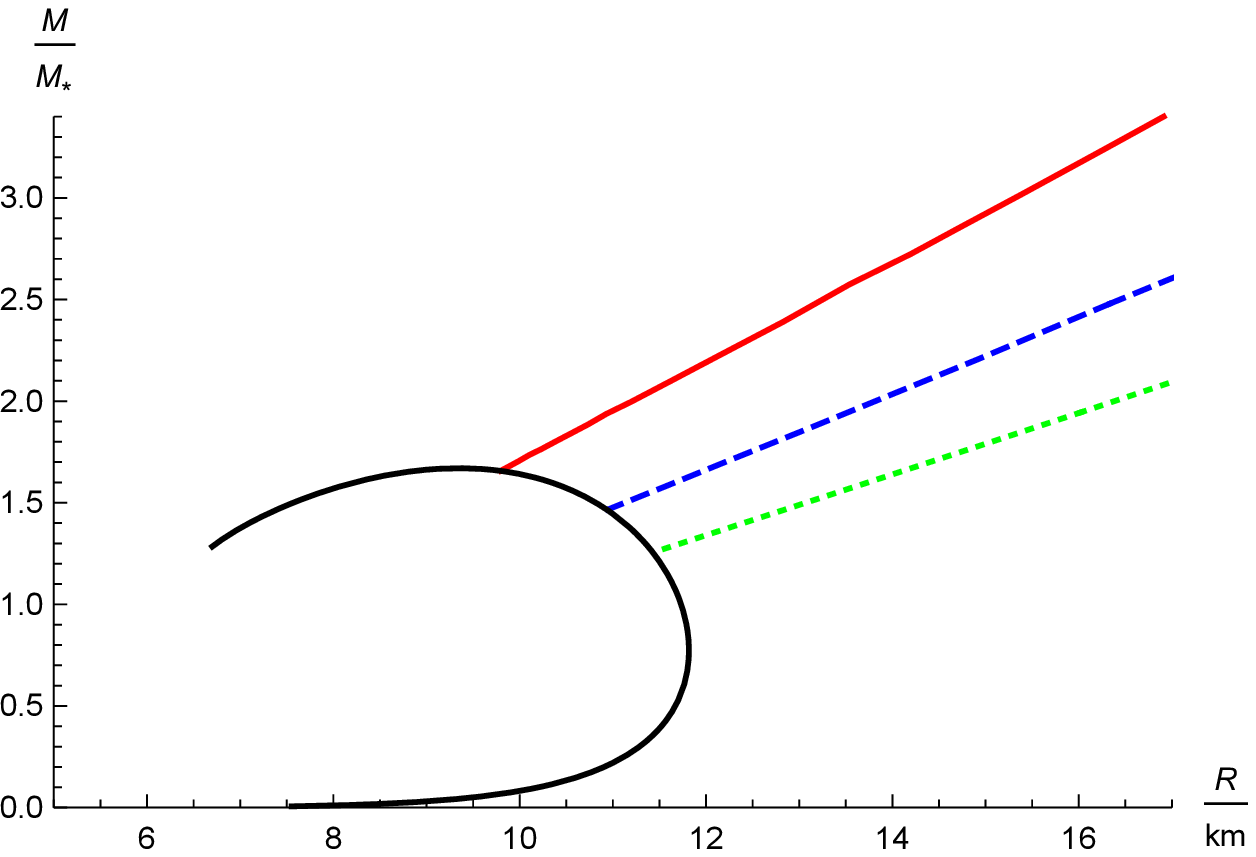}  
  \includegraphics[height=5.1cm,angle=0]{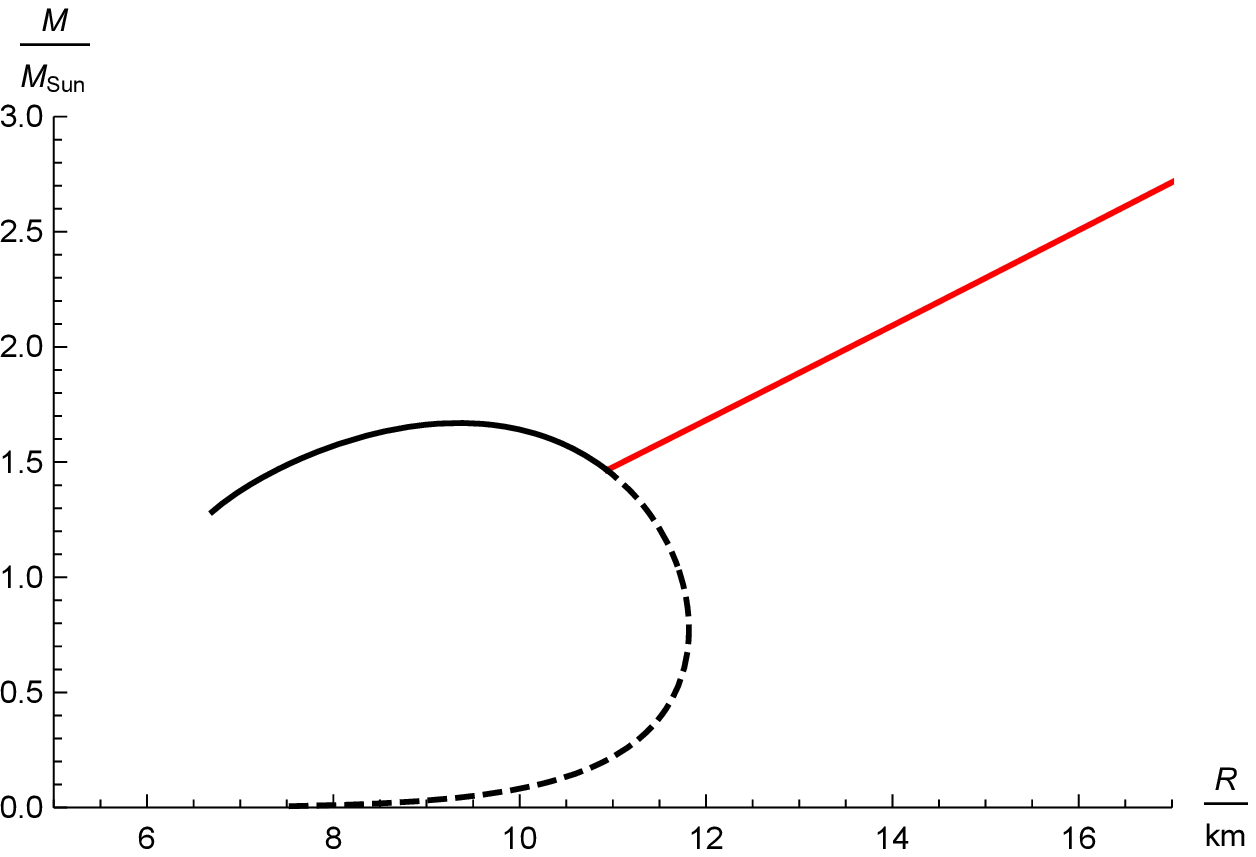}
\caption{
The mass-radius diagram is shown for the $0$ node solutions.
In the top panel,
the red, blue-dashed, and green-dotted curves 
correspond to
the cases of $B_0=-0.3,-0.5,-0.7$,
respectively,
while the black curve corresponds to the GR solutions.
The bottom panel focuses on the case of $B_0=-0.5$.
In the bottom panel, 
the black-dashed curve represents
the range of the branch of the GR solutions
for which the branch of the $0$ node solutions coexists.
The radius and mass are measured 
in the units of ${\rm km}$ and
the Solar mass $M_\odot$, respectively.
}
  \label{fig0node}
\end{center}
\end{figure} 
As $j$, i.e., ${\tilde p}_C$, decreases,
both the mass and radius increase.
On the other hand, 
as $j$ increases,
the branch of the nontrivial vector field solutions finally converges
to that of the GR solutions.

For $j={\cal O} (1.0)$,
the $0$ node solutions
exist only for $B_0={\cal O} (-0.1)$.
The similar behavior was also observed in the generalized Proca theories \cite{Kase:2020yhw}.
In our model, 
as suggested in Sec.~\ref{sec22},
the $0$ node solutions arise from the selected choice of the initial conditions
rather than the instability of the GR solutions,
as the hyperbolicity of the perturbations would be broken
on the GR stellar backgrounds.

\subsection{The $1$ node solutions}

Next, we focus on the $1$ node solutions.
In Fig. \ref{1node}, 
the typical behavior for the $1$ node solutions is shown for $B_0=-5.0$.
In both the panels,
the solid and dashed curves correspond to the cases of $j=1.0$ and $j=2.0$, respectively.
In the top panel,
the red and blue curves
represent
the energy density ${\tilde \rho}c^2$ and pressure ${\tilde p}$ inside the star,
respectively,
and the horizontal and vertical axes are shown 
in the units of ${\rm cm}$ and ${\rm dyne}\cdot {\rm cm}^{-2}$,
respectively.
The bottom panel shows
the temporal component of the vector field $A_t$
as the function of $r/{\cal R}$.
In the bottom panel,
the red points correspond to the surface of the star.
\begin{figure}[h]
\unitlength=1.1mm
\begin{center}
  \includegraphics[height=4.5cm,angle=0]{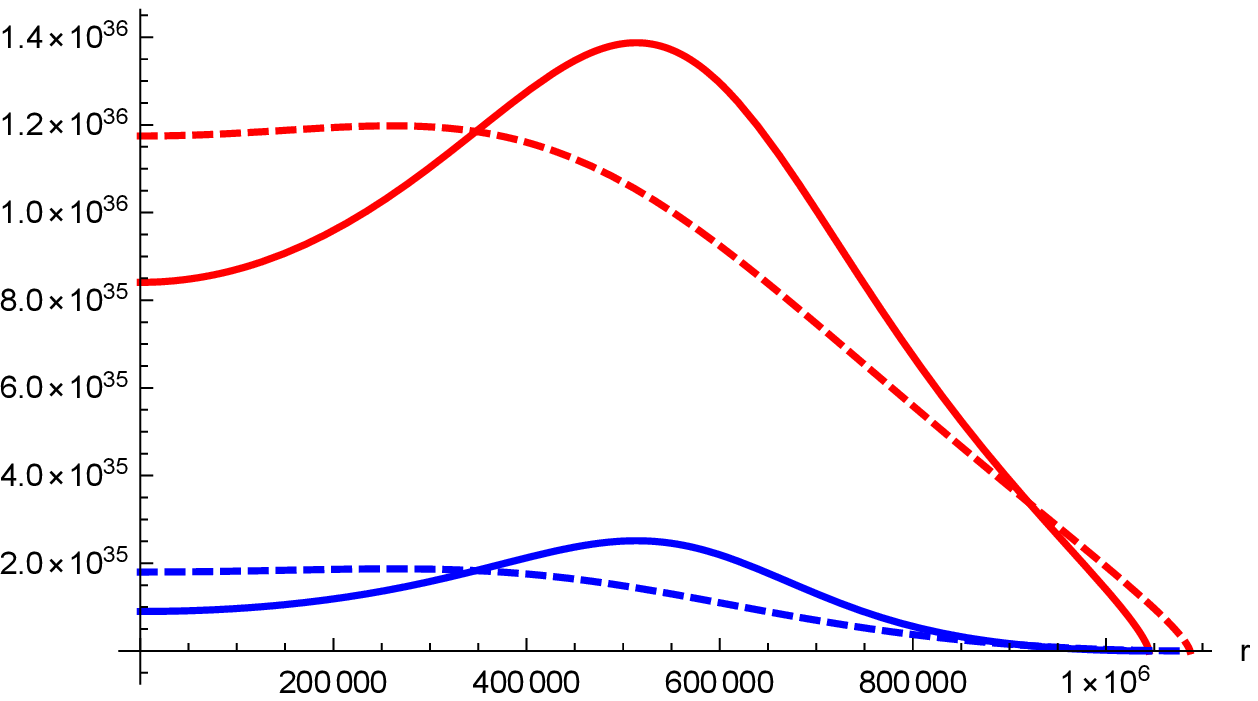}  
  \includegraphics[height=4.5cm,angle=0]{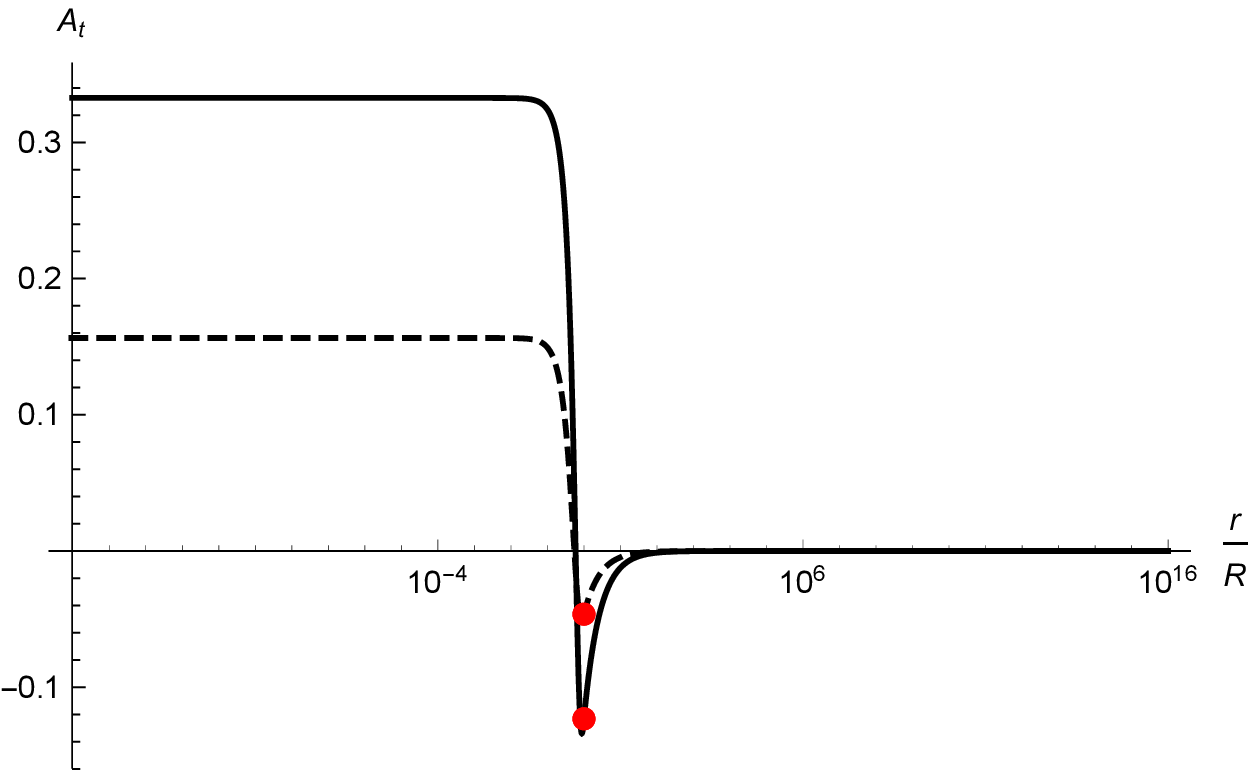}
\caption{
The typical behavior for the $1$ node solutions is shown for $B_0=-5.0$.
In both the panels,
the solid and dashed curves correspond to the cases of $j=1.0$ and $j=2.0$, respectively.
In the top panel,
the red and blue curves
represent
the energy density ${\tilde \rho}c^2$ and pressure ${\tilde p}$ inside the star,
respectively,
and the horizontal and vertical axes are shown 
in the units of ${\rm cm}$ and ${\rm dyne}\cdot {\rm cm}^{-2}$,
respectively.
The bottom panel shows
the temporal component of the vector field $A_t$
as the function of $r/{\cal R}$.
In the bottom panel,
the red points correspond to the surface of the star.
}
  \label{1node}
\end{center}
\end{figure} 

In Fig.~\ref{figroot1node},
$P$ is shown as the function of $A_C$ for $B_0=-5.0$.
The red-dashed, blue-dotted, green-dotdashed, and black curves correspond to
correspond to
the cases of $j=0.3, 1.0, 2.0, 3.0$,
respectively.
\begin{figure}[h]
\unitlength=1.1mm
\begin{center}
  \includegraphics[height=5.1cm,angle=0]{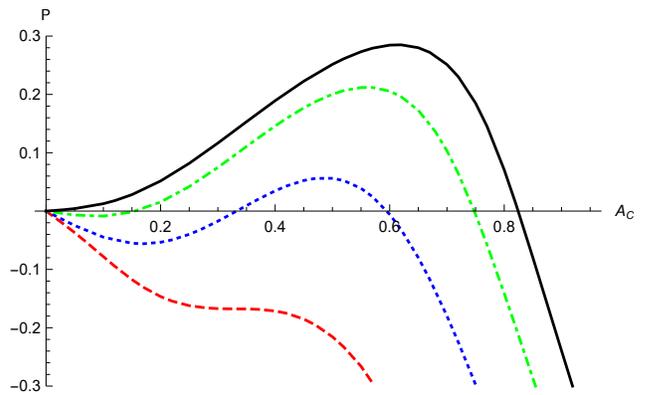}
\caption{
$P$ is shown as the function of  $A_C$ for the $0$ node solutions. 
The red-dashed, blue-dotted, green-dotdashed, and black curves correspond to
the cases of $j=0.5, 1.0, 2.0, 3.0$, respectively.
}
  \label{figroot1node}
\end{center}
\end{figure} 
For the smaller $j$, i.e.,  ${\tilde p}_C$,
as shown by the red curve, 
there is no root satisfying $P=0$,
which corresponds to the $1$ node solutions.
For the  larger $j$, i.e.,  ${\tilde p}_C$,
as shown by the blue and green curves,
two roots of $P=0$ appear at the finite $A_C$.
As $j$ further increases, 
the smaller root approaches $0$ and finally converges to the GR solutions $A_C=P=0$,
while the larger root keeps increasing.
For the sufficient large $j$, 
only the larger root exists.
As in Ref. \cite{Kase:2020yhw},
we focus on the smaller root and
show the mass-radius diagram.

In Fig. \ref{fig1node}, 
the top panel shows
the mass-radius diagram for the $1$ node solutions.
The red, blue-dashed, and green-dotted curves
correspond to
the cases of $B_0=-5.0,-6.0,-7.0$,
while the black curve corresponds to the GR solutions.
The bottom panel shows the solution for $B_0=-5.0$.
In the bottom panel, 
the black-dashed curve represents 
the range of the branch of the GR solutions
for which the branch of the $1$ node solutions coexists.
The radius and mass are measured 
in the units of ${\rm km}$ and
the Solar mass $M_\odot$,
respectively.
\begin{figure}[h]
\unitlength=1.1mm
\begin{center}
\includegraphics[height=5.1cm,angle=0]{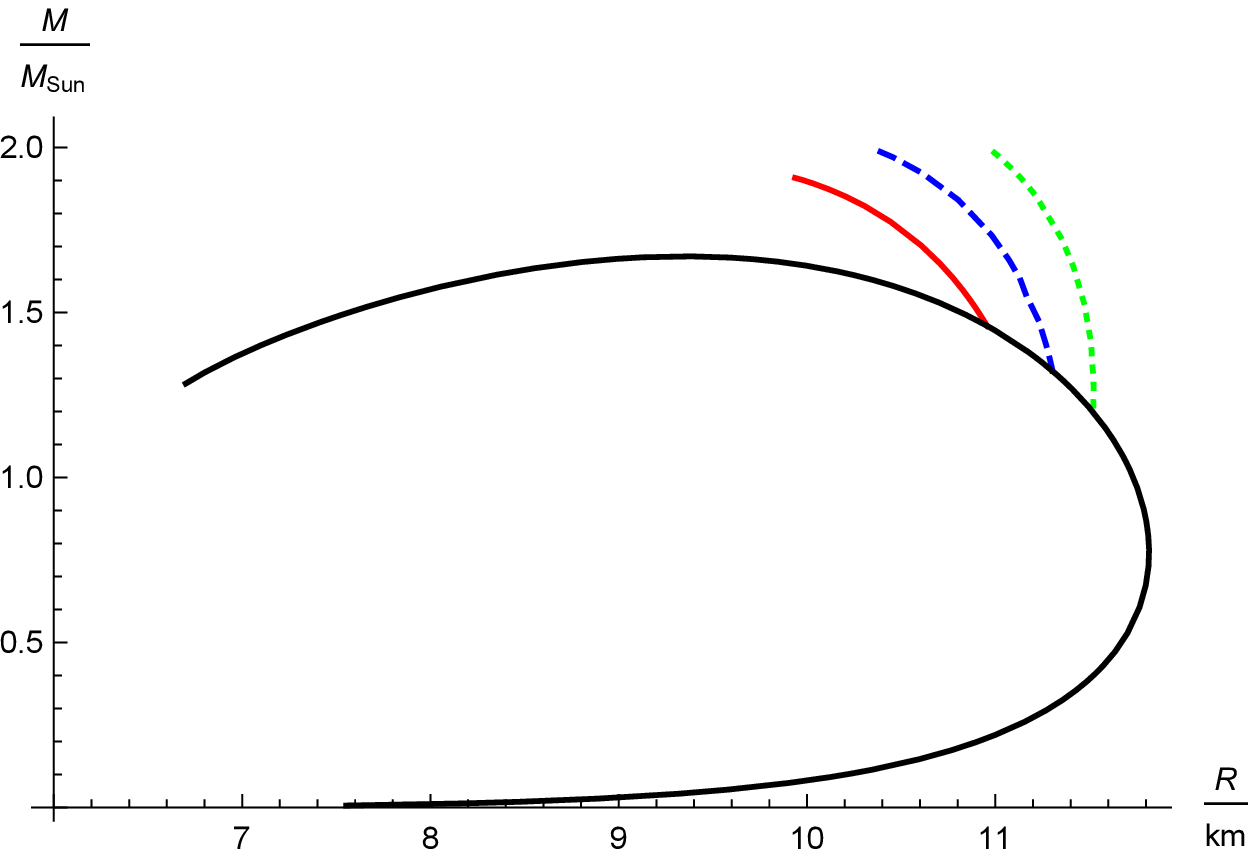}  
\includegraphics[height=5.1cm,angle=0]{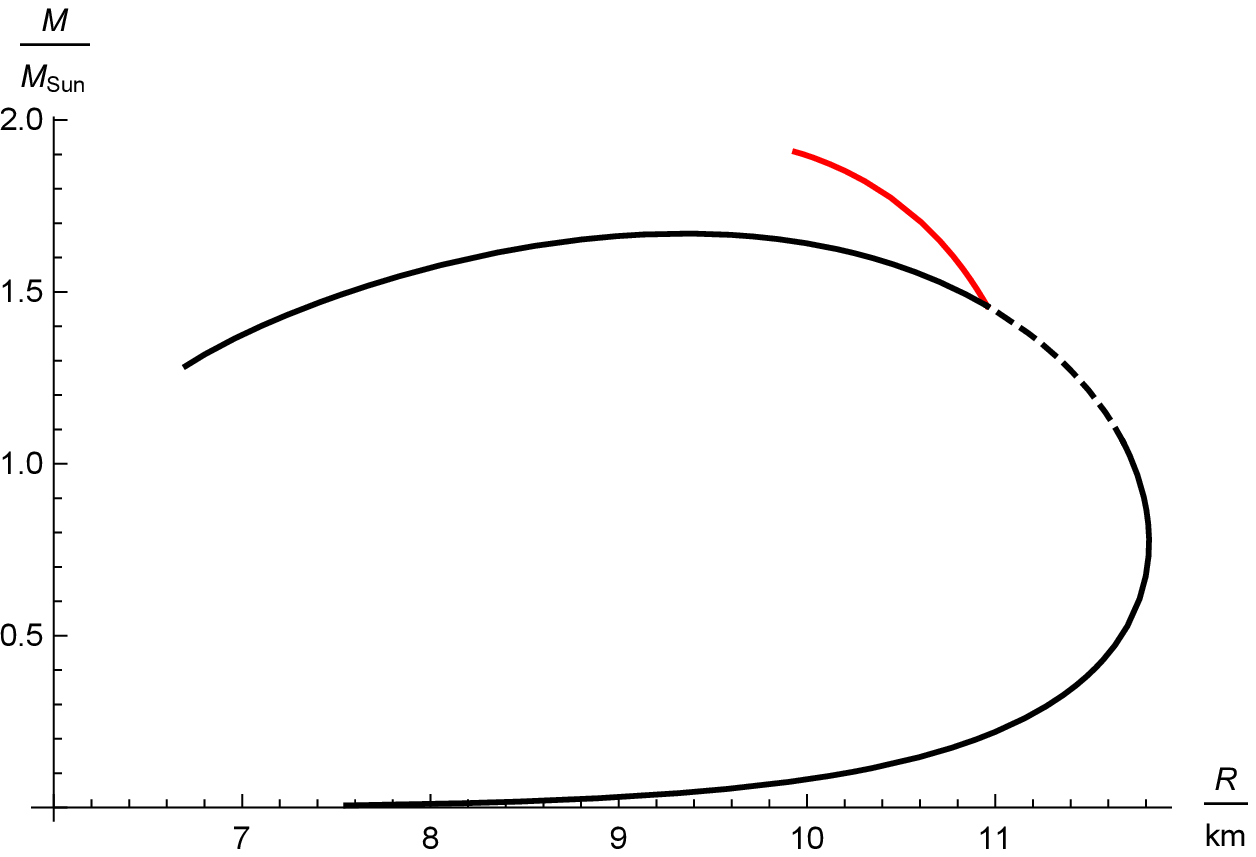}
\caption{
The top panel shows
the mass-radius diagram for the $1$ node solutions.
The red, blue-dashed, and green-dotted curves 
correspond to
the cases of $B_0=-5.0,-6.0,-7.0$,
while the black curve corresponds to the GR solutions.
The bottom panel shows the solution for $B_0=-5.0$.
In the bottom panel, 
the black-dashed curve represents 
the range of the branch of the GR solutions
for which the branch of the $1$ node solutions coexists.
The radius and mass are measured 
in the units of ${\rm km}$ and
the Solar mass $M_\odot$,
respectively.
}
  \label{fig1node}
\end{center}
\end{figure} 
In the bottom panel, 
for the smaller values of $j$,
i.e., smaller ${\tilde p}_C$,
the solution is described by the GR solution with the polytrope equation of state,
and across the critical central pressure $j\approx 0.785$,
it suddenly jumps to the nontrivial $1$ node solutions described by the red curve.
As $j$ further increases,
the radius increases but the mass decreases,
and 
the branch of the $1$ node solutions
finally converges to that of the GR solutions  
at $j\approx 2.63$.

For $j={\cal O} (1.0)$,
we observe that 
the $1$ node solutions exist for $B_0={\cal O} (-1.0)$.
The behavior of the $1$ node solutions
is also similar to what was observed in Ref. \cite{Kase:2020yhw}.
As seen in Fig. \ref{fig1node},
in the low density regimes
the branch of the $1$ node solutions is disconnected to the branch of the GR solutions.
This indicates that 
the $1$ node solutions 
may also be formed from the selected choice of the initial conditions.
Although we have not employed the more realistic equations of state, 
we expect that 
the essential results are insensitive to the choice of the equations of state.

\section{The case of the pure conformal coupling}
\label{sec6}

We then consider the model with the pure conformal coupling given by  
\begin{eqnarray}
\label{pure_conformal}
C(X)=1+C_1 X, 
\qquad 
B(X)=0,
\end{eqnarray}
where $C_1$ is the constant.
The relation of the mass and radius of the star in both the frames
is given by Eqs. \eqref{radius} and \eqref{mass}, respectively.
We then focus on the solutions satisfying the boundary condition $P=0$.
As we have seen in Sec. \ref{sec22},
these solutions may be formed
as the consequence of the tachyonic instability of the GR solution.
In this section,
we focus only on the $0$ node solutions,
as they can be obtained for $C_1={\cal O} (-1.0)$,
which is the value suggested from the analysis of the weak gravity limit
in Sec \ref{sec4}.
While the model with the massive vector field $m\neq 0$
was analyzed in Ref. \cite{Ramazanoglu:2017xbl},
we will focus on the model of the massless vector field $m=0$.
To our knowledge,
there has been no quantitative study
even in the massless case $m=0$ yet.

In Fig. \ref{conf},
the typical behavior for the $0$ node solutions is shown for $C_1=-1.0$.
In both the panels,
the solid and dashed curves correspond to the cases of $j=1.0$ and $j=2.0$, respectively.
In the top panel,
the red and blue curves
represent
the energy density ${\tilde \rho}c^2$ and pressure ${\tilde p}$ inside the star,
respectively,
and the horizontal and vertical axes are shown 
in the units of ${\rm cm}$ and ${\rm dyne}\cdot {\rm cm}^{-2}$,
respectively.
The bottom panel shows
the temporal component of the vector field $A_t$
as the function of $r/{\cal R}$.
In the bottom panel,
the red points correspond to the surface of the star.
\begin{figure}[h]
\unitlength=1.1mm
\begin{center}
  \includegraphics[height=4.5cm,angle=0]{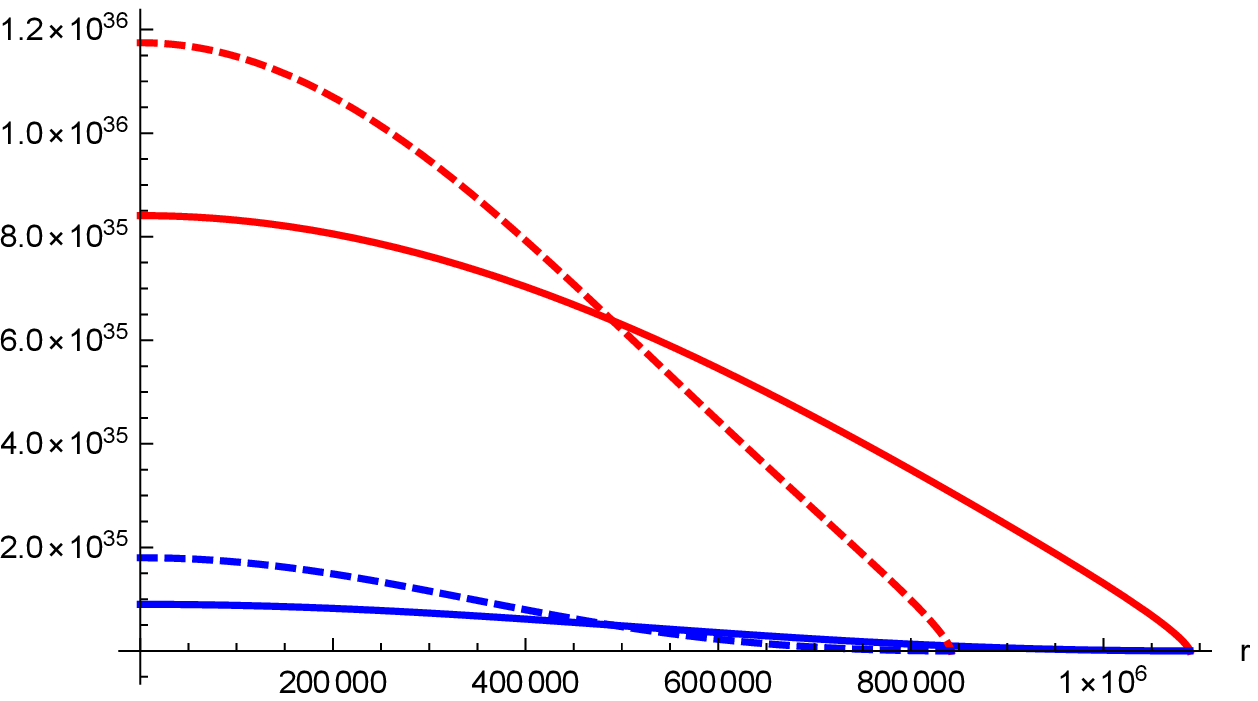}  
  \includegraphics[height=4.5cm,angle=0]{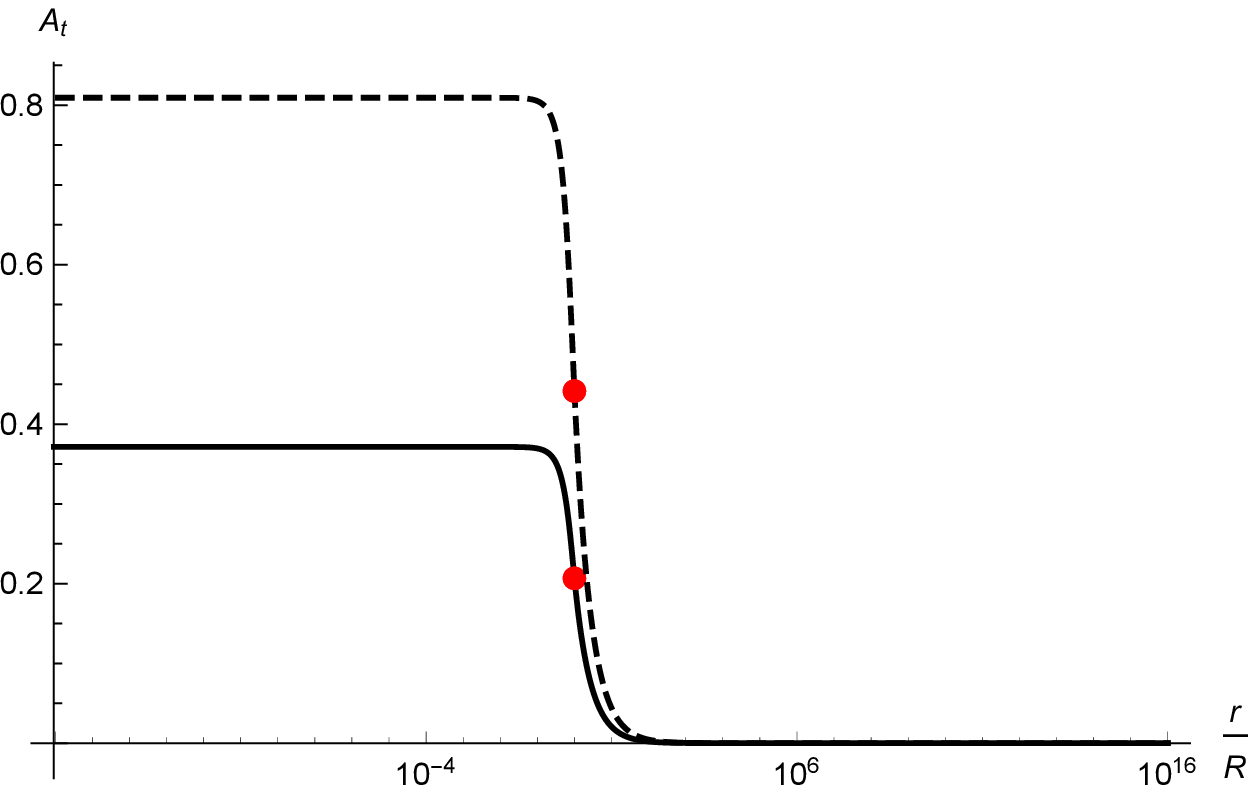}
\caption{
The typical behavior for the $0$ node solutions is shown for $C_1=-1.0$.
In both the panels,
the solid and dashed curves correspond to the cases of $j=1.0$ and $j=2.0$, respectively.
In the top panel,
the red and blue curves
represent
the energy density ${\tilde \rho}c^2$ and pressure ${\tilde p}$ inside the star,
respectively,
and the horizontal and vertical axes are shown 
in the units of ${\rm cm}$ and ${\rm dyne}\cdot {\rm cm}^{-2}$,
respectively.
The bottom panel shows
the temporal component of the vector field $A_t$
as the function of $r/{\cal R}$.
In the bottom panel,
the red points correspond to the surface of the star.
}
  \label{conf}
\end{center}
\end{figure} 

In Fig.~\ref{figconfdata1},
following Eq. \eqref{charges},
$P$ is shown as the function of $A_C$ 
for $C_1=-1.0$ (the top panel)
and $C_1=-2.0$ (the bottom panel). 
In the top panel,
the black, red-dashed, blue-dotted, and green-dotdashed curves
correspond to
the cases of $j= 0.5, 1.0, 1.5, 2.0$,
respectively.
In the bottom panel,
the black, red-dashed, blue-dotted, and green-dotdashed curves 
correspond to
the cases of $j=0.01, 0.05, 0.1, 0.5$,
respectively.
\begin{figure}[h]
\unitlength=1.1mm
\begin{center}
  \includegraphics[height=5.1cm,angle=0]{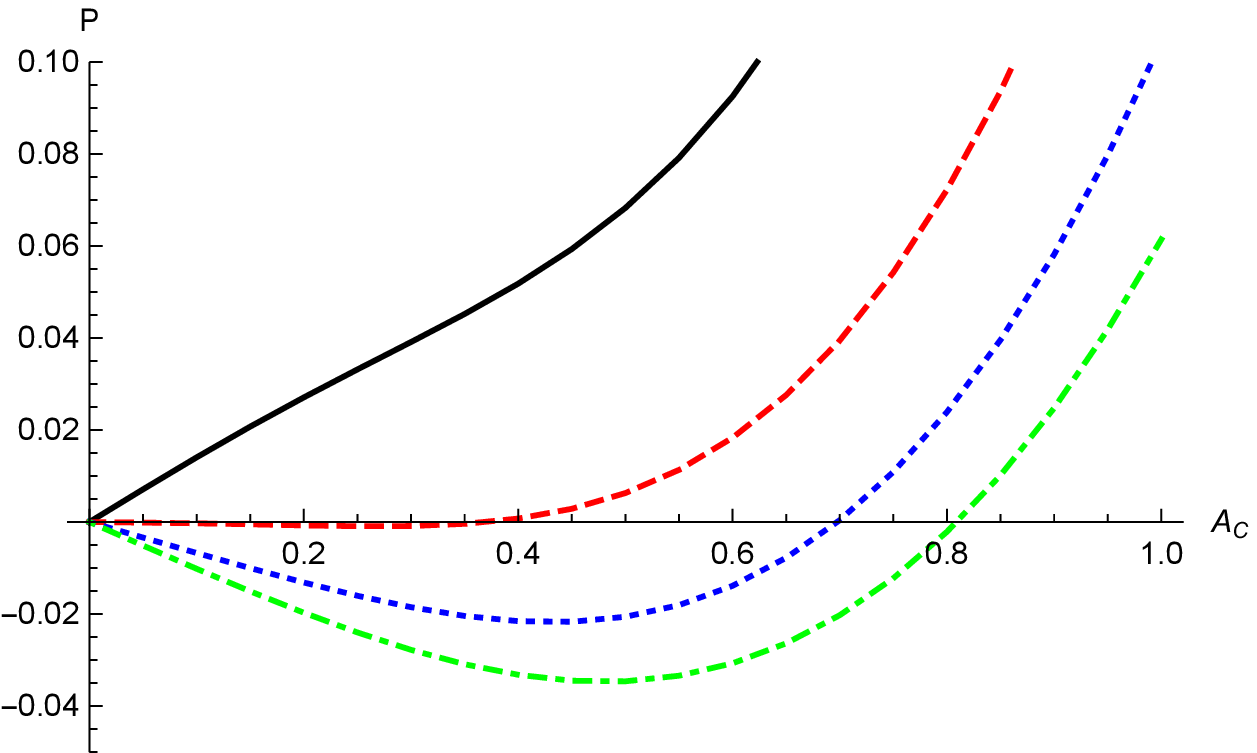}  
  \includegraphics[height=5.1cm,angle=0]{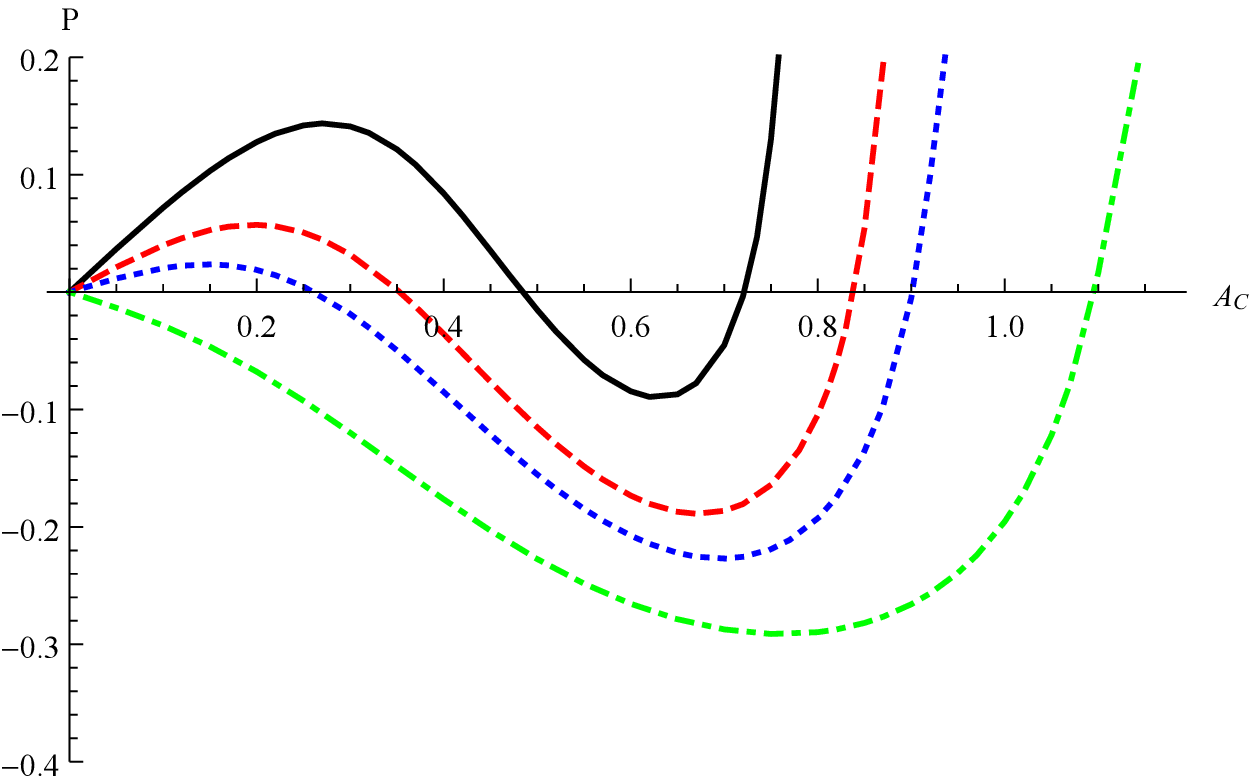}
\caption{
$P$ is shown as the function of $A_C$ 
for $C_1=-1.0$ (the top panel)
and $C_1=-2.0$ (the bottom panel). 
In the top panel,
the black, red-dashed, blue-dotted, and green-dotdashed curves 
correspond to
the cases of $j= 0.5, 1.0, 1.5, 2.0$,
respectively.
In the bottom panel,
the black, red-dashed, blue-dotted, and green-dotdashed curves
correspond to
the cases of $j=0.01, 0.05, 0.1, 0.5$,
respectively.
}
  \label{figconfdata1}
\end{center}
\end{figure} 
In both the panels,
all curves take $P=0$ at $A_C=0$,
which correspond to the GR solutions.
For $C_1=-1.0$,
the curves of $j=1.0, 1.5, 2.0$ cross the axis of $P=0$ at $A_C\neq 0$ 
which corresponds to the nontrivial $0$ node solutions,
while 
the curve of $j=0.5$
has no root other than $A_C=0$.
As $j$, i.e., ${\tilde p}_C$, increases,
the root of $A_C\neq 0$ increases,
then decreases
and
finally converges
to the GR solutions $A_C=0$. 
Thus, the nontrivial $0$ node solutions 
exist in the intermediate range of ${\tilde p}_C$,
and 
in the low and high density regimes
they converge to the GR solutions.

On the other hand,
for $C_1=-2.0$,
the curve of $j=0.5$
crosses the axis of $P=0$ once at $A_C= {\cal O}(1.0)$.
The curves for $j=0.01, 0.05, 0.1$
cross the axis of $P=0$ twice.
The smaller root
exists only for $j<{\cal O} (0.1)$,
and
at the certain $j$
converges to the branch of the GR solutions.
The larger root keep increasing 
for the increasing $j$.
Thus,
the behavior for $C_1=-2.0$
is qualitatively different from the cases of $|C_1|\gtrsim {\cal O}(2.0)$,
and especially 
the low density regimes
the $0$-node solutions
are not smoothly connected to the GR solutions.

In Fig.~\ref{mrconf},
the mass-radius diagram for the $0$ node solutions is shown.
The red, blue-dashed, green-dotted, and purple-dotdashed curves
correspond to
the cases of the $0$-node solutions for $C_1= -1.0, -1.2, -1.5, -2.0$,
respectively,
while the black curve corresponds to the case of the GR solutions.
The radius and mass are measured 
in the units of ${\rm km}$ and
the Solar mass $M_\odot$, respectively.
\begin{figure}[h]
\unitlength=1.1mm
\begin{center}
  \includegraphics[height=5.1cm,angle=0]{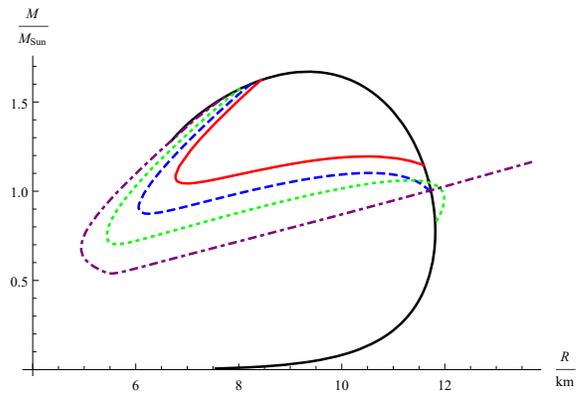}
\caption{
The mass-radius diagram for the $0$ node solutions is shown.
The red, blue-dashed, green-dotted, and purple-dotdashed curves 
correspond to
the cases of $C_1= -1.0, -1.2, -1.5, -2.0$,
respectively,
while the black curve corresponds to the GR solutions.
The radius and mass are measured 
in the units of ${\rm km}$ and
the Solar mass $M_\odot$,
respectively.
}
  \label{mrconf}
\end{center}
\end{figure} 
For $C_1=-1.0,-1.2,-1.5$,
the branch of $0$ node solutions
is connected to the branch of the GR solutions. 
Thus,
as in the case of spontaneous scalarization, 
in the mass-radius diagram,
the branch of the $0$ node solutions
bifurcates from the branch of the GR solutions in the low density regimes.
After the significant deviations from the branch of the GR solutions,
they finally merge to the GR solutions in the high density regimes. 
For the same radius of the star,
the mass of the star for $0$ node solutions 
becomes smaller than that for the GR solutions.
On the other hand, 
for $C_1=-2.0$
the branch of the $0$-node solutions
is disconnected to the branch of the GR solutions in the low density regime,
while it marges to the branch of the GR solutions in the high density regimes.
For such a case,
spontaneous vectorization would not happen 
in the same manner as spontaneous scalarization.
We note that our study is the first one 
which shows the mass-radius diagram
for the vectorized relativistic stars
in the case of the massless vector-tensor theories.

In Fig. \ref{mq},
the dimensionless ratio $Q/M$ is shown as the function of the mass $M$.
The black and red-dashed curves 
correspond to
the cases of $C_1= -1.0, -1.2$,
respectively.
The mass is measured in the units of the Solar mass $M_\odot$.
\begin{figure}[h]
\unitlength=1.1mm
\begin{center}
  \includegraphics[height=5.1cm,angle=0]{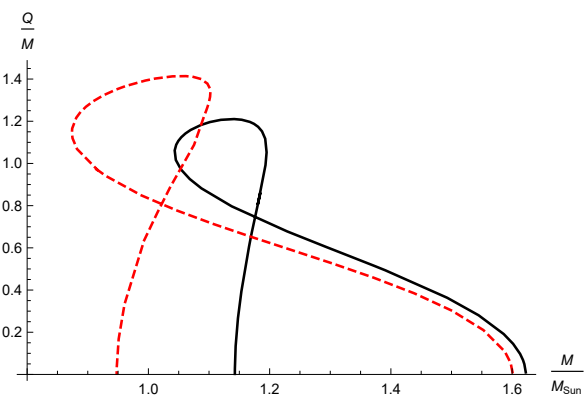}
\caption{
The dimensionless ratio $Q/M$ is shown as the function of the mass.
The black and red-dashed curves 
correspond to
the cases of $C_1= -1.0, -1.2$,
respectively.
The mass is measured in the units of the Solar mass $M_\odot$.
}
  \label{mq}
\end{center}
\end{figure} 
Since vectorized solutions with two different masses
can share the same radius, 
in some range of the mass
the dimensionless ratio $Q/M$ also becomes 
the multivalued function of $M$.
The dimensionless ratio $Q/M$ exceeds unity
around its maximal value
and approaches zero at the both endpoints,
as the vectorized branch approaches the GR branch
in the low and high density regimes.
The maximal value of $Q/M$ also increases
as $|C_1|$ increases.

Although
we have focused on the simplest model \eqref{pure_conformal},
we expect that 
spontaneous vectorization would still take place
for the more general couplings \eqref{regular_coupling} with $C_1<0$ and $C_n\neq 0$ ($n=2,3,\cdots$).
While
the behavior in the vicinity of the bifurcation points from the branch of the GR solutions 
in the low and high density regimes in the mass-radius diagram 
would remain the same,
the behavior in the intermediate density regime
would depend crucially on the choice of the coupling function $C(X)$.
The case of the other coupling function $C(X)$
and the inclusion of the nonzero bare mass of the vector field
 $m\neq 0$ will be left for the future work.

\section{The case of the combined conformal and disformal couplings}
\label{sec7}

Before closing this paper, 
we briefly discuss the case 
of the combined effects of the conformal and disformal couplings
on the nontrivial solutions.
For simplicity,
we consider the minimal extension of the models 
considered in Secs. \ref{sec5} and \ref{sec6},
\begin{eqnarray}
\label{conformal_disformal}
C(X)=1+C_1 X, 
\qquad 
B(X)=B_0,
\end{eqnarray}
For a fixed nonzero value of $C_1={\cal O} (-1.0)$
where we have obtained the vectorized solutions with $0$ nodes for $B_0=0$
in Sec. \ref{sec6},
we will see how adding the nonzero value of $B_0$
modifies the structure of the nontrivial solutions
and the existence of the solutions.

In Fig, \ref{figcombined},
$P$ is shown as the function of $A_C$ 
for $C_1=-1.0$ and $j=1.5$.
In the top panel, 
the black, red-dashed, blue-dotted, and green-dotdashed curves
correspond to the cases
of $B_0=0, -0.1, -0.15, -0.2$, respectively.
In the bottom panel, 
the black, red-dashed, blue-dotted, and green-dotdashed curves
correspond to the cases
of $B_0=-0.3, -1.0, -1.5, -2.0$, respectively.
\begin{figure}[h]
\unitlength=1.1mm
\begin{center}
  \includegraphics[height=5.1cm,angle=0]{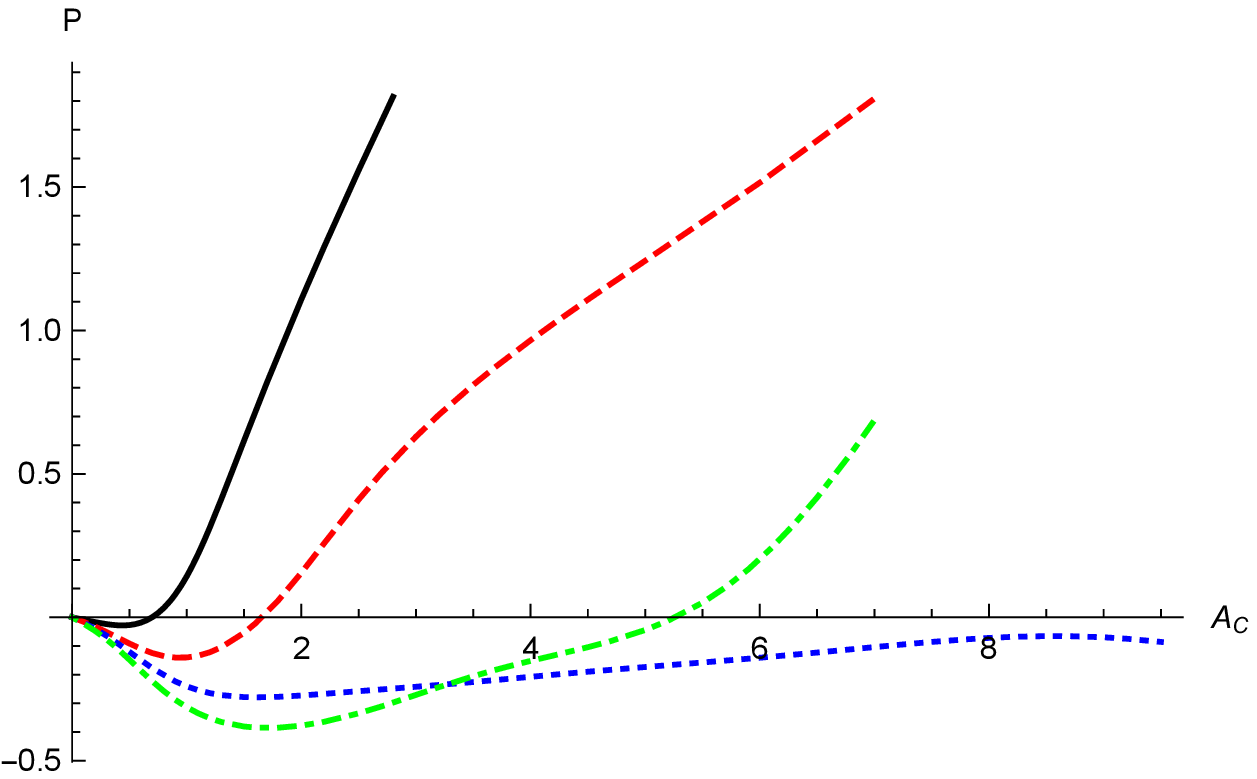}  
  \includegraphics[height=5.1cm,angle=0]{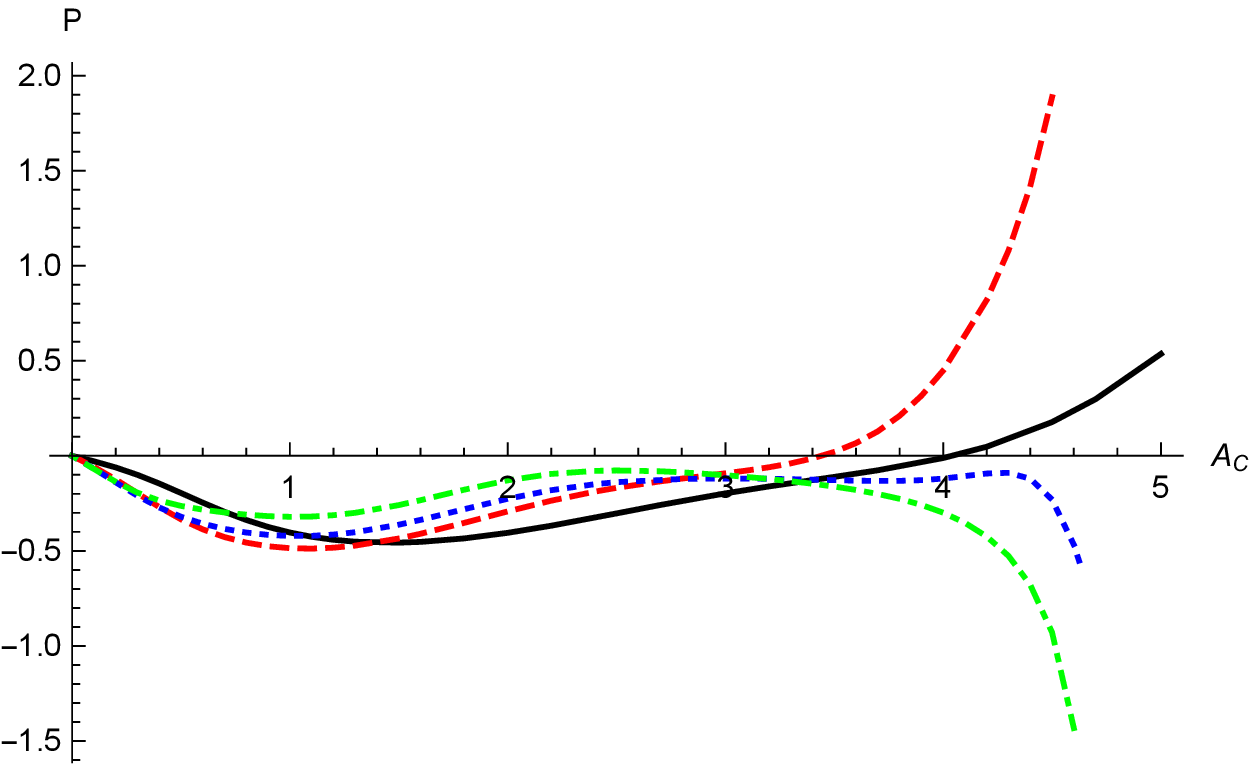}
\caption{
$P$ is shown as the function of $A_C$ 
for $C_1=-1.0$ and $j=1.5$.
In the top panel, 
the black, red-dashed, blue-dotted, and green-dotdashed curves
correspond to the cases
of $B_0=0, -0.1, -0.15, -0.2$, respectively.
In the bottom panel, 
the black, red-dashed, blue-dotted, and green-dotdashed curves
correspond to the cases
of $B_0= -0.3, -1.0, -1.5, -2.0$, respectively.
}
  \label{figcombined}
\end{center}
\end{figure} 
For $B_0=-0.1$,
the solution of $P=0$ corresponds to the solution with $0$ nodes
as for $B_0=0$.
For $B_0=-0.15$,
there is no solution of $P=0$.
For $B_0=-0.2$ and $-0.3$,
there are the 0-node solutions,
but as seen in Fig. \ref{comb},
the profile of the vector field possesses
a spike inside the star.
\begin{figure}[h]
\unitlength=1.1mm
\begin{center}
  \includegraphics[height=4.5cm,angle=0]{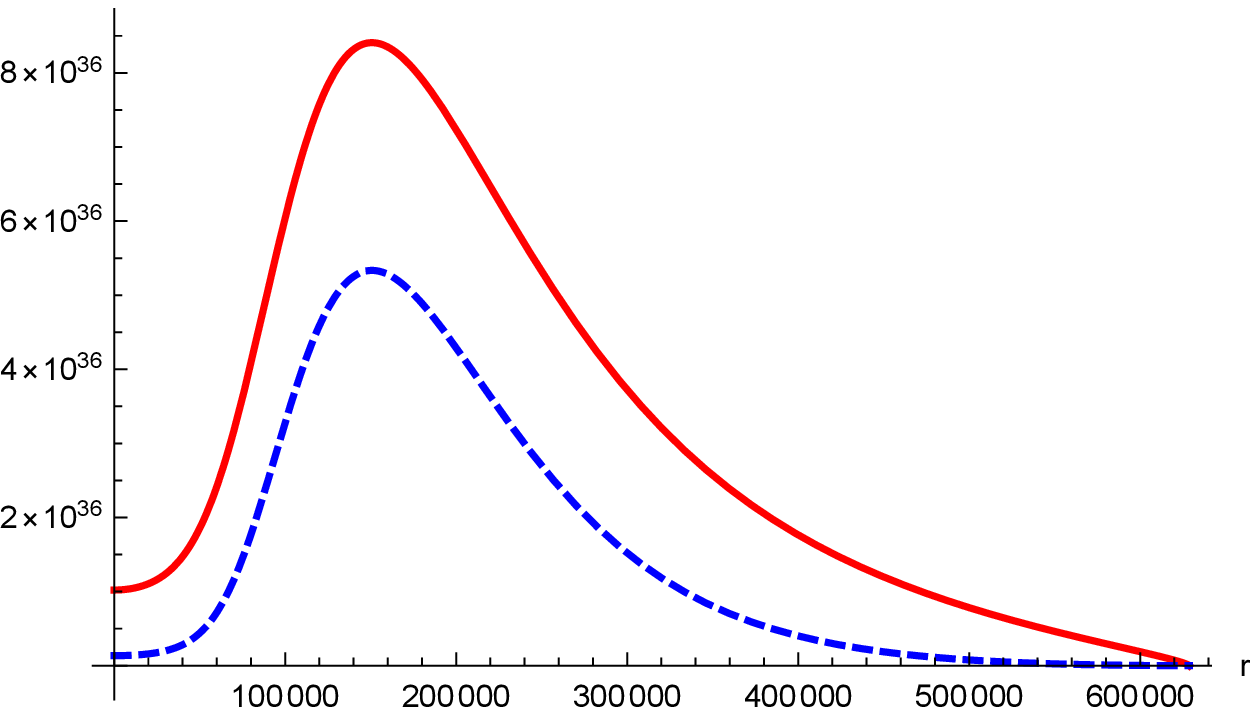}  
  \includegraphics[height=4.5cm,angle=0]{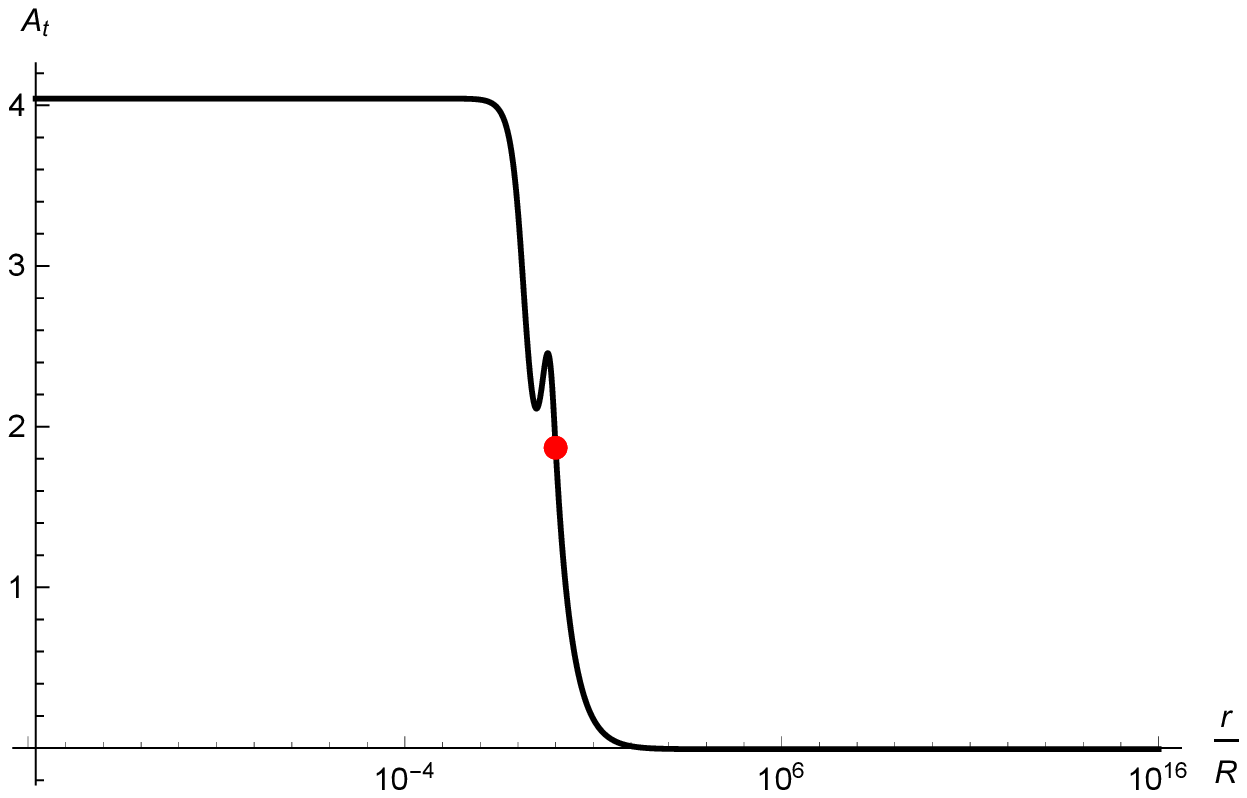}
\caption{
The $0$ node solution is shown for $C_1=-1.0$, $B_0=-0.3$,
and $j=1.5$.
In the top panel,
the red and blue-dashed curves
represent
the energy density ${\tilde \rho}c^2$ and pressure ${\tilde p}$ inside the star,
respectively,
and 
the horizontal and vertical axes are shown in the units of ${\rm cm}$
and ${\rm dyne}\cdot {\rm cm}^{-2}$, respectively.
The bottom panel shows
the temporal component of the vector field $A_t$
as the function of $r/{\cal R}$.
In the bottom panel,
the red point corresponds to the surface of the star.
}
  \label{comb}
\end{center}
\end{figure} 
The solution of  $P=0$ for $B_0=-1.0$
corresponds to a $1$-node solution 
with a single spike inside the star.
For these spiky solutions,
as in the top panel of Fig. \ref{comb},
both the energy density and pressure increase
in the vicinity of the center of the star.

Thus, 
adding the disformal coupling to the pure conformal coupling 
makes the properties of the nontrivial solutions 
more unphysical than in the case of the pure conformal coupling.
In other words,
the model with the pure conformal coupling 
would be the most competitive one for spontaneous vectorization.

\section{Conclusions}
\label{sec8}

In this paper, 
we have examined the possibility of spontaneous vectorization in the vector-tensor theories 
with the vector conformal and disformal couplings to matter
given by the action Eq. \eqref{action} with \eqref{disformal}.
In each class of the couplings, 
we have constructed the static and spherically symmetric solutions of the relativistic stars 
with the nontrivial profile of the vector field satisfying the boundary conditions at the spatial infinity, $A_\mu (r\to \infty)=0$.
We have worked from the viewpoint of the Einstein frame,
and note that 
the gravitational action \eqref{action} rewritten in terms of the Jordan frame metric 
belongs to a class of the beyond-generalized Proca theories.

First, 
we have considered the general couplings which
allow the general relativistic (GR) solutions.
After varying the action,
we have derived the covariant equations of motion.
Then,
applying to the static and spherically symmetric spacetime,
we have derived the modified Tolman-Oppenheimer-Volkoff equations.
The radial component of the vector field equation
forces us to set the radial component of the vector field to zero, $A_r=0$.
In the case that the bare mass of the vector field is zero,
the exterior solution is given by the Reissner-Nordstr\"om solution.
The external solution is matched to the internal solution
constructed by solving the modified Tolman-Oppenheimer-Volkoff equations
from the center to the surface of the star
under the regularity boundary conditions at the center of the star.
The matching at the surface of the star  
determines the mass and the vector field charge of the star.

After deriving the covariant equations of motion,
we have investigated the linear perturbations about the static and spherically symmetric GR solutions.
Focusing on the radial perturbations, 
we have found that 
the vector conformal and disformal couplings cause different kinds of instability.
In the case of the pure conformal coupling, 
we have shown 
that the GR solutions suffer from the tachyonic instability,
indicating that
the GR solutions would spontaneously evolve to the nontrivial solutions
with the nonzero value of the vector field.

On the other hand, 
in the case of the pure vector disformal coupling,
we argued that the GR solutions suffer
from the ghost or gradient instability,
indicating the breakdown of the hyperbolicity on the GR stellar backgrounds.
Thus, 
we have expected that 
the pure disformal coupling would not cause 
spontaneous vectorization
in the same manner as spontaneous scalarization,
and the stellar solutions with the nontrivial profile of the vector field
may be formed from the selected class of the initial conditions.

Second,
we have focused on the simplest model with the pure disformal coupling~\eqref{pure_disformal}.
We have constructed the static and spherically symmetric solutions with the asymptotic value $P=0$.
The properties of the stellar solutions are similar to those 
in the generalized Proca theories~\cite{Kase:2020yhw}.
The $0$ node solutions
exist for the coupling parameter of ${\cal O} (-0.1)$,
while the $1$ node solutions exist for the coupling parameter of ${\cal O} (-1.0)$.
As expected,
in the mass-radius diagram the branch of the $0$ node solutions 
is disconnected to that of the GR solutions in the low density regimes.

On the other hand, 
we have also studied the simplest model with the pure conformal coupling~\eqref{pure_conformal}.
We have argued that the $0$ node solutions are formed 
spontaneously via the continuous evolution from the GR solutions.
For the smaller absolute value of the conformal coupling parameter,
we have found that the branch of the $0$ node solutions
is connected to that of the GR solutions in the low density regimes,
and hence
these $0$ node solutions may be the vectorized solutions
arising from the tachyonic instability of the GR solutions.
The mass of the $0$ solutions is always smaller than that of the GR solutions.
On the other hand, 
for the larger absolute value of the coupling parameter,
the branch of the $0$ node solutions
is disconnected to that of the GR solutions in the low density regimes,
and hence
the $0$ node solutions may be formed
from the selected choice of the initial conditions.

We have also seen the behavior of the nontrivial
solution in the presence of both the conformal and disformal couplings.
Although we could obtain the nontrivial vector field solutions
satisfying $P=0$ in this model,
adding the disformal coupling to the pure conformal coupling model
has made the properties of the stars more unphysical than
in the case of the pure conformal coupling,
as a spiky feature appears in the profile of the vector field
and both the energy density and pressure increase
in the vicinity of the center of the star.

While in the present work
we have focused on the simplest cases
of the pure vector conformal and disformal couplings
with the polytrope equation of state,
the analysis of the more general models
with the more realistic equations of state
will be left for the future studies.
We will also leave the analysis of
the time evolution
toward the nontrivial solutions of the vector field for the future work.
Clarifying these points
will make the difference of the properties of spontaneous vectorization
from those of spontaneous scalarization
more evident.

\acknowledgments{
M.M.~was supported by the research grant under the Decree-Law 57/2016 of August 29 (Portugal) through the Funda\c{c}\~{a}o para a Ci\^encia e a Tecnologia, and CENTRA through the Project~No.~UIDB/00099/2020.
}

\appendix

\section{Some useful formulas}

The inverse and determinant of the metric \eqref{disformal}
are given by 
\begin{eqnarray}
\label{app1}
&&
{\tilde g}^{\mu\nu}
=
\frac{1}{C}
\left[
g^{\mu\nu}
-\frac{B}{1+BX}
A^\mu A^\nu
\right],
\nonumber\\
&&
{\tilde g}=C^4 g\left(1+BX \right).
\end{eqnarray}


The nonzero components of the metric in the Jordan frame
in the static and spherically symmetric system
are given by
\begin{eqnarray}
\label{jordan_sss}
&&
{\tilde g}_{tt}= C(-e^{\nu}+B A_t^2),
\qquad
{\tilde g}_{tr}=BC A_t A_r,
\nonumber\\
&&
{\tilde g}_{rr}= C(e^\lambda  +B A_r^2),
\qquad
{\tilde g}_{ab} =C  r^2 \gamma_{ab}.
\end{eqnarray}
The components of the matter energy-momentum tensor in the Einstein frame
in the static and spherically symmetric system
are given by
\begin{widetext}
\begin{eqnarray}
\label{em_sss}
&&
T^t{}_t
=
-\frac{C^2}{\sqrt{1+BX}}
\left\{
\left(
 1+ e^{-\lambda}B A_r^2
 \right)
 \tilde\rho
+ e^{-\nu}A_t^2
\left[
B_X
\left(
e^{-\nu} {\tilde \rho} A_t^2
+
e^{-\lambda} {\tilde p}_r A_r^2
\right)
+\frac{C_X}{C}
(1+BX)
\left(
-{\tilde \rho}
+{\tilde p}_r
+2{\tilde p}_t
\right) 
\right]
\right\},
\nonumber\\
&&
T^t{}_r
=
-\frac{C^2 e^{-\nu}A_tA_r}{\sqrt{1+BX}}
\left\{
B \tilde\rho
+B_X
\left(
e^{-\nu} {\tilde \rho} A_t^2
+
e^{-\lambda} {\tilde p}_r A_r^2
\right)
+\frac{C_X}{C}
(1+BX)
\left(
-{\tilde \rho}
+{\tilde p}_r
+2{\tilde p}_t
\right)
\right\},
\nonumber\\
&&
T^r{}_t
=\frac{C^2 e^{-\lambda}A_r A_t}{\sqrt{1+BX}}
\left\{
-B {\tilde p}_r
+B_X
\left(
e^{-\nu} {\tilde \rho} A_t^2
+
e^{-\lambda} {\tilde p}_r A_r^2
\right)
+\frac{C_X}{C}
(1+BX)
\left(
-{\tilde \rho}
+{\tilde p}_r
+2{\tilde p}_t
\right)
\right\},
\nonumber\\
&&
T^r{}_r
=\frac{C^2}{\sqrt{1+BX}}
\left\{
\left(
1-e^{-\nu}B A_t^2
 \right)
{\tilde p}_r
+
e^{-\lambda}A_r^2
\left[
B_X
\left(
e^{-\nu} {\tilde \rho} A_t^2
+
e^{-\lambda} {\tilde p}_r A_r^2
\right)
+\frac{C_X}{C}
(1+BX)
\left(
-{\tilde \rho}
+{\tilde p}_r
+2{\tilde p}_t
\right)
\right]
\right\},
\nonumber\\
&&
T^a{}_b
=\sqrt{1+BX} C^2
{\tilde p}_t
\delta^a{}_b.
\end{eqnarray}
\end{widetext}
When $A_r=0$, 
it reduces to the perfect fluid form as 
\begin{eqnarray}
\label{perfect_eins}
&&
\rho
:=-T^t{}_t
=
\frac{C^2}{\sqrt{1+BX}}
\nonumber\\
&\times&
\left[
 \tilde\rho
\left(
1+B_X X^2
\right)
-
\frac{C_X X}{C}
(1+BX)
\left(
-{\tilde \rho}
+{\tilde p}_r
+2{\tilde p}_t
\right) 
\right],
\nonumber\\
&&
p_r:=
T^r{}_r
=C^2
\sqrt{1+BX}
{\tilde p}_r,
\nonumber\\
&&
p_t:= 
\frac{1}{2}
T^a{}_a
=
C^2
\sqrt{1+BX} 
{\tilde p}_t.
\end{eqnarray}
The coefficients in the approximated solutions 
in the vicinity of the center of the star Eq. \eqref{center}
are given by 
\begin{widetext}
\begin{eqnarray}
\label{central}
\mu_3
&=&
\frac{1}{24\sqrt{1+BX_0}}
\left\{
-m^2 X_0 \sqrt{1+ BX_0}
+
4{\tilde \rho}_0 \kappa^2 C^2 
\left(
1+X_0^2 B_X
\right)
+4\kappa^2    C C_X X_0
 ({\tilde \rho}_0-3{\tilde p}_0)
  (1+ BX_0)
\right\},
\nonumber\\
\nu_2
&=&
\frac{1}{6\sqrt{1+ BX_0}}
\left\{
-m^2X_0 \sqrt{1+ BX_0}
+ \kappa^2 C^2 
\left(
{\tilde \rho}_0
+3{\tilde p}_0
+3 {\tilde p}_0 B X_0
+ {\tilde\rho}_0 B_X X_0^2
\right)
+\kappa^2 CC_X X_0
({\tilde \rho}_0
-3{\tilde p}_0)
(1+BX_0)
\right\},
\nonumber\\
\frac{a_2}{A_C}
&=&
\frac{m^2}{6}
+\frac{\kappa^2 C^2}{3\sqrt{1+BX_0}}
\left[
{\tilde \rho}_0 C
\left(B+B_X X_0\right)
+\left({\tilde\rho_0}-3{\tilde p}_0\right)C_X
 \left(1+B X_0\right)
\right],
\nonumber\\
{\tilde p}_2
&=&
\frac{X_0 B_X}{2(1+BX_0)}
\left[
 2e^{-\nu_0/2} a_2 ({\tilde\rho}_0+{\tilde p}_0)\sqrt{-X_0}
+ \nu_2 ({\tilde p}_0-{\tilde \rho}_0)X_0
\right]
+\frac{C_X}{2C}
\left[
 2 e^{-\nu_0/2} a_2 ({\tilde\rho}_0+{\tilde p}_0)\sqrt{-X_0}
+ \nu_2 (7{\tilde p}_0-{\tilde \rho}_0) X_0 
\right]
\nonumber\\
&-&
\frac{{\tilde\rho}_0+{\tilde p}_0}{2(1+BX_0)}
\left(
\nu_2-2e^{-\nu_0/2}  a_2 B \sqrt{-X_0} 
\right).
\end{eqnarray}
\end{widetext}

\bibliographystyle{apsrev4-1}
\bibliography{disformal_refs}

\end{document}